\documentclass{scrartcl}

\usepackage{amsthm,amssymb,amsmath,bm,graphicx,graphbox,
  dsfont,xcolor,float,booktabs,hyperref,units,mathtools,subcaption}
\usepackage[Algorithm]{algorithm}
\usepackage{setspace}
\usepackage{placeins}
\usepackage{algpseudocode,mycommands,tikz}
\usetikzlibrary{shapes.geometric,positioning}
\theoremstyle{definition}

\theoremstyle{remark}
\newtheorem*{remark}{Remark}
\begin{document}

\newcommand{\cl}[1]{{\color{blue} \noindent CL: #1}}
\newcommand{\red}[1]{{\color{red} \noindent #1}}
\newcommand{\blue}[1]{{\color{blue} \noindent #1}}



\title{A neural network multigrid solver for the Navier-Stokes equations}
\author{Nils Margenberg\thanks{
    Helmut Schmidt University,
    Holstenhofweg 85, 22043 Hamburg,
    Germany,
    \texttt{margenbn@hsu-hh.de}
  }
  \and Dirk Hartmann\thanks{
    Siemens AG,
    Corporate Technology,
    Otto-Hahn-Ring 6,
    81739 Munich, Germany,
    \texttt{hartmann.dirk@siemens.com}
  }
  \and Christian Lessig\thanks{
    University of Magdeburg,
    Institute for Simulation and Graphics,
    Universit\"atsplatz 2,
    39104 Magdeburg,
    Germany,
    \texttt{christian.lessig@ovgu.de}
  }
  \and Thomas Richter\thanks{
    University of Magdeburg,
    Institute for Analysis and Numerics,
    Universit\"atsplatz 2,
    39104 Magdeburg,
    Germany,
    \texttt{thomas.richter@ovgu.de}}
}

\maketitle

\begin{abstract}
  We present the deep neural network multigrid solver (DNN-MG) that we develop for the instationary Navier-Stokes equations.
  DNN-MG improves computational efficiency using a judicious combination of a geometric multigrid solver and a recurrent neural network with memory.
  DNN-MG uses the multi-grid method to classically solve on coarse levels while the neural network corrects interpolated solutions on fine ones, thus avoiding the increasingly expensive computations that would have to be performed there.
  This results in a reduction in computation time through DNN-MG's highly compact neural network.
  The compactness results from its design for local patches and the available coarse multigrid solutions that provides a ``guide'' for the corrections.
 A compact neural network  with a small number of parameters also reduces training time and data.
Furthermore, the network's locality facilitates generalizability and allows one to use DNN-MG trained on one mesh domain also on different ones.
 We demonstrate the efficacy of DNN-MG for variations of the 2D laminar flow around an obstacle.
 For these, our method significantly improves the solutions as well as lift and drag functionals while requiring only about half the computation time of a full multigrid solution.
 We also show that DNN-MG trained for the configuration with one obstacle can be generalized to other time dependent problems that can be solved efficiently using a geometric multigrid method.
\end{abstract}

\section{Introduction}
\label{sec:intro}

The last decade has seen enormous progress on the use of deep neural networks for applications in machine translation, computer vision, and many other fields, e.g.~\cite{LeCun2015,Schrittwieser2019}.
These advances lead to a growing (and renewed) interest to apply neural networks also to problems in computational science and engineering, including for the simulation of partial differential equations.
There, the existence of efficient numerical methods raises the questions how neural networks can be combined with, e.g. multigrid finite element methods, to leverage the benefits the different methodologies provide.

Towards this end, we propose the deep neural network multigrid solver (DNN-MG) that we develop in this paper for the simulation of the instationary Navier-Stokes equations.
DNN-MG combines a geometric multigrid solver with a recurrent neural network to replace the computations on one or multiple finest mesh levels with a network-based correction.  
For this, the Navier-Stokes equations are first solved on the first $L$ levels of the multigrid mesh hierarchy using a classical multigrid solver. 
The obtained solution is then interpolated (or prolongated) to a finer level $L+K$ where the network predicts a correction of the velocity vector field towards the unknown ground truth.
The right hand side of the problem is then formed on the fine level $L+K$ and its restriction to level $L$ is used in the classical multigrid solve in the next time step.
Through this, the neural network-based corrections feed back to the coarse levels and affect the overall time evolution of the flow while the costs of the multigrid computations on fine levels are avoided. 

The key to the efficiency of the DNN-MG solver is the use of a highly compact neural network with a small number of parameters.
The compactness is achieved by the network's locality, i.e. that it operates independently on patches of the mesh domain, e.g. a single mesh element on level $L$ or a collection of few adjacent elements.
It is also enabled by the availability of the coarse multigrid solution that provides significant information about the flow behavior and thus reduces the state that needs to be carried in the network.
A local, patch-based network also reduces the required amount of training data and time, since even a single flow contains already a multitude of patches and different behaviors.
The latter also facilitates the networks' ability to generalize to flows not seen during training.
To ensure that the neural network can model complex flow behavior, it contains a Gated Recurrent Unit (GRU) with memory that incorporates a flow's past into a prediction, similar to how many classical time integration schemes use multiple past time steps for their update.
The GRU's memory and the use of the network's prediction in the right hand side at time step $n+1$ facilitates a stable and temporally consistent time integration.

We demonstrate the efficacy of the DNN-MG solver for a 2D laminar flow in a channel with obstacles with varying elliptical eccentricities. 
We first consider the classical flow with one obstacle and train with different eccentricities than used for testing.
The obtained solutions as well as the lift and drag functionals demonstrate that DNN-MG obtains considerably better accuracy than a coarse multigrid solution on level $L$ while requiring only about half the computation time of a full solution on level $L+1$.
For a prediction using level $L+2$ even better results are obtained with only a very small additional overhead compared to DNN-MG on level $L+1$. 
To analyze DNN-MG's ability to generalize to flow regime's not seen during training, we also consider the channel flow with no or with two obstacles while using the neural network trained for the one obstacle case.
In both instances, DNN-MG is able to provide high fidelity solutions and for the two obstacle flow it again reduces the error of the lift and drag functionals substantially.

The remainder of the paper is structured as follows.
In the next section we review related work that uses neural networks for the simulation of partial differential equations.
In Sec.~\ref{sec:fe} we then recall the solution of the Navier-Stokes equations using the geometric multigrid method before providing an introduction to recurrent neural networks in Sec.~\ref{sec:nn}.
In Sec.~\ref{sec:dnnmg} we introduce the deep neural network multigrid solver, discuss its design and also sketch its generalization to other problems.
Our numerical results are presented in Sec.~\ref{sec:num}.

\section{Related work}

Recently, there has been an increasing interest to employ (deep) neural networks for the simulation of physical systems.
In the following, we will discuss existing approaches with an emphasis on those that consider partial differential equations and in particular the Navier-Stokes equations.

For learning, partial differential equations are often considered as sequential time series.
Neural network architectures developed for these data can then be employed.
The most common architecture are recurrent neural networks (RNN), e.g. with Long-Short-Term-Memory units (LSTM)~\cite{Hochreiter1997} to avoid the vanishing gradient problem. 
Because of the prevalence of sequential data, LSTMs led to many variations over the years.
One are Gated Recurrent Units (GRUs) which, while not as powerful as LSTMs, provide often similar performance and are easier to train~\cite{Cho2014}.
An alternative to LSTM-type architectures are temporal convolutional neural networks (TCNs) where the temporal dependence is modeled using a causal (i.e. one-sided) convolution in the time variable.
Bai et al.~\cite{Bai2018} demonstrated that this can provide substantial improvements in prediction accuracy over recurrent neural networks with memory.
A different approach to modeling temporal dependencies was developed by Voelker et al.~\cite{Voelker2019} who represented the time domain in Legendre polynomials and learned basis function coefficients.
Through this, they obtained a neural network model with continuous time.
A related idea are the neural ordinary differential equations architectures by Chen et al.~\cite{Chen2018} where the depth of the network is continuous.

Approaches for a direct description of the time evolution of partial differential equations using neural networks go back more than 20 years, e.g.~\cite{Lagaris1998}.
In recent work, Raissi et al.~\cite{Raissi2018b} demonstrated that the velocity field and pressure of the Navier-Stokes equations can be learned to good approximation only from observing passively advected tracer particles. 
Nabian and Meidani~\cite{Nabian2018}, Yang and Perdikaris~\cite{Yang2018}, and Raissi and Karniadakis~\cite{Raissi2018} exploited that not only observations are available but also a known analytic model.
These authors hence also use the deviation of the network prediction from the model as an additional penalty term in the training loss.
Kasim et al.~\cite{Kasim2020} recently demonstrated neural network-based simulations for a broad range of applications, including fluid dynamics, by also optimizing the network architecture itself during the training process.
Eichinger, Heinlein and Klawonn~\cite{EichingerHeinleinKlawonn2020} use convolutional neural networks and techniques from image processing to learn flow patterns for the Navier-Stokes flow around objects of different shape. 

Next to the above direct neural network-based approaches to the simulation of partial differential equations, there have been different attempts to integrate neural networks into existing numerical techniques.
As in our work, the objective is to combine the benefits of neural networks and classical methods and obtain techniques that would be difficult or impossible with either approach alone.
For elasticity, neural network-based representations for constitutive relations were learned in ~\cite{Tartakovsky2018,Berg2019,Rudy2019}.
These were then used in classical simulations, e.g. based on finite elements.
Wiewel et al.~\cite{Wiewel2018} presented a simulation of the Navier-Stokes equation where an LSTM-based deep neural network is used to predict the pressure correction. 
Xi et al.~\cite{Xie2018} and Werhahn et al.~\cite{Werhahn2019} use neural networks to upsample a coarse simulation and obtain physically plausible high frequency details.
An approach based on the splitting method for high dimensional PDEs was developed by Beck et al.~\cite{Beck2019} where one obtains a set of smaller learning problems that are easier to treat than a single large one.
Wan and Sapsis~\cite{Wan2018} describe the motion of finite size particles in fluid flows by combining analytic arguments with a neural network that models the discrepancy between the (highly) idealized analytic model and real world observations.

To ensure that neural network-based predictions respect physical invariants, such as energy conservation or divergence freedom, network architectures tailored towards physical simulations have been proposed in the literature.
The authors of~\cite{Mattheakis2019,Chen2019,Jin2020} developed neural networks that incorporate the symplectic structure of Hamiltonian mechanics and they demonstrate that this improves generalization and accuracy.
In a related line of research, ideas from nonlinear analysis and stability theory were exploited.
Erichson et al.~\cite{Erichson2019}, for example, improve the accuracy of the learning process by ensuring that the learned system has the same Lyapunov stability as the system of interest. 
Similarly, Gin et al.~\cite{Gin2019} train a neural network to provide a coordinate transformation that linearizes a nonlinear PDE, in the spirit of the Cole-Hopf transform for Burgers equation, but using a neural network to find and represent the transform. 
For this, they use a Koopman operator representation.
Related to this is the work Zhang et al.~\cite{Zhang2019} where a neural network is used to improve the dynamically orthogonal (DO) and bi-orthogonal (BO) methods for stochastic pdes.

An alternative line of research aims to learn surrogate neural network representations of physical systems from the governing (stochastic) partial differential equations without recourse to simulations or measurements~\cite{Geneva2020,Zhu2019,Karumuri2020}.
This avoids the training data generation that is often a bottleneck with the approaches above and allows one to quantify the uncertainty of the prediction.
Karumuri et al.~\cite{Karumuri2020}, e.g., obtain their loss function through a variational formulation of the stochastic PDEs they are considering.
Related to this is the Deep Ritz Method by E and Yu~\cite{E2018} where the ansatz of the Ritz method is used but with neural networks for the function approximation.
Qin et al.~\cite{Qin2019} show that residual networks can be seen as exact one-step time stepping methods and these provide building blocks for multi-step methods.



The performance of (deep) neural networks has also been analyzed theoretically.
Already in the 1980s and 1990s, the first results demonstrated the potential of neural networks as universal function approximators~\cite{Cybenko1989,Hornik1991} that do not suffer from the curse of dimensionality~\cite{Barron1994}.
More recent results on the approximation properties of deep neural networks focused in particular on the required size of a network to achieve certain approximation bound, e.g.~\cite{Bolcskei2019}.
Mallat~\cite{Mallat2016} provided insight into the effectiveness of deep neural networks using tools from harmonic analysis and pointed out the importance of invariants for their understanding.
For partial differential equations, currently only few results exist. 
Kutyniok et al.~\cite{Kutyniok2019} studied the theoretical power of forward, deep neural networks for reduced order models and they show that the solution map, which yields the basis coefficients of the reduced basis for given values of the parameters, can be efficiently approximated using neural networks and in a manner essentially independent of the dimension.
Grohs et al.~\cite{Grohs2019} study the spacetime error for PDEs when Euler steps are realized using neural networks.
To our knowledge, for partial differential equations no theoretical results for recurrent neural networks with memory exist.

\section{Finite element discretization of the Navier-Stokes equations}
\label{sec:fe}

In the following, we provide an overview of the solution of the Navier-Stokes equations using the geometric multigrid method.

We consider a domain $\Omega\in \R^d$ with $d\in \{2,\,3\}$ and
Lipschitz-continuous boundary $\Gamma$ and a bounded time interval
$[0,\,T]$. The solution of the
instationary Navier-Stokes equations then seeks the
velocity $v\colon [0,\,T]\times \Omega \to \R^d$ and pressure
$p\colon [0,\,T]\times \Omega \to \R$, such that
\begin{equation}
  \begin{alignedat}{2}\label{eq:nsstrong}
    \partial_t v + (v\cdot \nabla)v - \frac{1}{\mathrm{Re}}\Delta v
    +\nabla p &= f \quad &&\text{on } [0,\,T] \times \Omega\\
    \nabla \cdot v &= 0 \quad &&\text{on } [0,\,T] \times \Omega,
  \end{alignedat}
\end{equation}
where $\mathrm{Re}>0$ is the Reynolds number and $f$ an external force.
The initial and boundary conditions are given by
\begin{equation}\label{eq:boundary}
  \begin{alignedat}{2}
    v(0,\,\cdot) &= v_0(\cdot)\quad &&\text{on }\Omega\\
    v &= v^D \quad &&\text{on } [0,\,T] \times \Gamma^D\\
    \frac{1}{Re}(\vec n\cdot\nabla)v - p\vec n &=0 \quad &&\text{in } [0,\,T] \times \Gamma^N ,
  \end{alignedat}
\end{equation}
where $\vec n$ denotes the outward facing unit normal vector on the boundary $\partial\Omega$ of the domain.
The boundary $\Gamma = \Gamma^D \cup \Gamma^N$ is split into subsets $\Gamma^D$ with Dirichlet boundary conditions and $\Gamma^N$ with Neumann type conditions.

\subsection{Variational formulation of the Navier-Stokes equations}

Eq.~\ref{eq:nsstrong} can be discretized with finite elements.
We briefly sketch the variational formulation in the following.
By $L^2(\Omega)$ we denote the space of square integrable functions on the domain $\Omega\subset\mathds{R}^d$ with scalar product $(\cdot,\cdot)$ and by $H^1(\Omega)$ those $L^2(\Omega)$ functions with weak first derivative in $L^2(\Omega)$.
The function spaces for the velocity and pressure are then
\begin{equation}\label{eq:VQ}
  \begin{alignedat}{1}
    V &\coloneqq v^D + H^1_0(\Omega;\Gamma^D)^d,\quad H_0^1 (\Omega;\Gamma^D)^d\coloneqq \brc{v\in H^1(\Omega)^d\colon v=0 \text{  on } \Gamma^D}\\
    L &\coloneqq \brc{p\in L^2(\Omega),\text{ and, if }\Gamma^N=\emptyset,\; \int_{\Omega}p \drv x = 0},
  \end{alignedat}
\end{equation}
where $v^D\in H^1(\Omega)^d$ is an extension of the Dirichlet data on $\Gamma^D$ into the domain. Given Dirichlet data on the complete boundary, the pressure is normalized to yield uniqueness.

With these spaces, the variational formulation of Eq.~\ref{eq:nsstrong} is given by
\begin{equation}
  \begin{alignedat}{2}\label{eq:ns}
    (\partial_t v,\,\phi) + (v\cdot \nabla v,\, \phi) +
    \frac{1}{\mathrm{Re}}(\nabla v,\, \nabla \phi) -
    (p,\,\nabla\cdot \phi)
    &= (f,\,\phi)\quad&&\forall \phi \in H^1_0(\Omega;\Gamma^D)^d\\
    (\nabla \cdot v, \, \xi) &= 0\quad &&\forall \xi \in L\\
    v(0,\,\cdot ) &= v_0(\cdot)\quad &&\text{on }\Omega.
  \end{alignedat}
\end{equation}
Let $\Omega_h$ be a quadrilateral or hexahedral finite element mesh of the domain $\Omega$ satisfying the usual requirements on the structural and form regularity such that the standard interpolation results hold, compare~\cite[Section 4.2]{Richter2017}.
For the finite element discretization of Eq.~\ref{eq:ns}, we choose $W_h^{(r)}$ as the space of continuous functions which are polynomials of degree $r$ on each element $T\in\Omega_h$.
If the finite element mesh contains elements of generalized quadrilateral or hexahedral type, e.g. to resolve curved boundaries, we use isoparametric finite elements,
see~\cite[Section 4.2.1]{Richter2017} for details and the specific realization chosen for this work.
The discrete trial- and test-spaces for the discretization of Eq.~\ref{eq:ns} we throughout the paper be $v_h,\,\phi_h \in V_h = [W_h^{(2)}]^d$ and $p_h,\,\xi_h \in L_h = W_h^{(2)}.$ Since the resulting equal order finite element pair $V_h\times L_h$ does not fulfill the inf-sup condition, we add additional stabilization terms of local projection type~\cite{BeckerBraack2001}. The resulting semidiscrete variational problem then reads
\begin{equation}
  \begin{alignedat}{2}\label{eq:disc:ns}
    (\partial_t v_h,\,\phi_h) + (v_h\cdot \nabla v_h,\, \phi_h) +
    \frac{1}{\mathrm{Re}}(\nabla v_h,\, \nabla \phi_h) -
    (p_h,\,\nabla\cdot \phi_h)
    &= (f,\,\phi_h)\quad&&\forall \phi_h\in V_h
    \\[5pt]
    (\nabla \cdot v_h, \, \xi_h)
    +\sum_{T\in\Omega_h}\alpha_T (\nabla (p_h-\pi_h p_h),\nabla (\xi_h-\pi_h \xi_h))
    &= 0\quad &&\forall \xi_h \in L_h
  \end{alignedat}
\end{equation}
where $\pi_h:W_h^{(2)}\to W_h^{(1)}$ denotes the local projection operator into the space of linear polynomials.
The local stabilization parameter $\alpha_T$ is chosen as
\[
\alpha_T = \alpha_0\cdot  \mathrm{Re} \cdot h_T^2,
\]
where we usually use $\alpha_0=0.1$ and where $h_T$ is the local mesh size of element $T$. We refer to~\cite{BeckerBraack2001,BraackBurmanJohnEtAl2007} for details. Since the numerical experiments described in Section~\ref{sec:num} consider moderate Reynolds numbers only, no stabilization of the convective term is required.


\subsection{Time discretization}
For temporal discretization, the time interval $[0,\,T]$ is split into discrete time steps of uniform size
\[
0=t_0<t_1<\cdots <t_N=T,\: k = t_n-t_{n-1}.
\]
The generalization to a non-equidistant time discretization is straightforward and only omitted  for ease of presentation.
We define $v_n\coloneqq v_h(t_n)$ and $p_n\coloneqq p_h(t_n)$ for the fully discrete approximation of velocity and pressure at time $t_n$ and apply the second order Crank-Nicolson method to Eq.~\ref{eq:disc:ns}, resulting in the fully discrete problem
\begin{align}
  \label{eq:4cranknicholson3}
  &(\nabla \cdot v_{n}, \, \xi_h)+
  \sum_{T\in\Omega_h}\alpha_T (\nabla (p_n-\pi_h p_n),\nabla (\xi_h-\pi_h \xi_h))
  =0 \qquad \forall \xi_h\in L_h, \nonumber \\[4pt]
  &\frac{1}{k}(v_n,\,\phi_h)\,
  +{}\frac{1}{2} (v_n\cdot \nabla v_n,\,\phi_h)
  +\frac{1}{2\mathrm{Re}}(\nabla v_n,\,\nabla \phi_h)
  -(p_n,\,\nabla \cdot \phi_h) \nonumber \\
  & \qquad =\frac{1}{k}(v_{n-1},\, \phi_h)
    +\frac{1}{2}(f_n,\phi_h)
    +\frac{1}{2}(f_{n-1},\,\phi_h) \nonumber \\
    & \qquad \quad \quad -\frac{1}{2}(v_{n-1}\cdot\nabla v_{n-1},\,\phi_h)
    - \frac{1}{2\mathrm{Re}}(\nabla v_{n-1},\,\nabla \phi_h)
    \quad \quad \quad \forall \phi_h\in V_h.
\end{align}
The right hand side only depends on the the velocity $v_{n-1}$ at the last time step $n-1$ and we will denote it as $b_{n-1}$ in the following.

The Crank-Nicolson time discretization in Eq.~\ref{eq:4cranknicholson3} corresponds to a nonlinear system of equations that can be solved, e.g., with the Newton-Krylov method.
In the presented form it is sufficiently robust for smooth initial data $v_0$; we refer to~\cite{HeywoodRannacher1990} for small modifications with improved robustness and stability.

%

\subsection{Newton-Krylov solution}
\label{sec:fe:newton_krylow}

In each time step, Eq.~\ref{eq:4cranknicholson3} amounts to a large nonlinear system of algebraic equations.
By introducing the unknown $x=(v_n,p_n)$, it can be written as
\begin{equation}\label{nonlinearshort}
  \begin{aligned}
    {\cal A}_h(x) &= f_h\\
    [{\cal A}_h(x)]_i &\coloneqq
    (\nabla \cdot v_{n}, \, \xi_h^i)+
    \sum_{T\in\Omega_h}\alpha_T (\nabla (p_n-\pi_h p_n),\nabla (\xi_h^i-\pi_h \xi_h^i))\\
    & \qquad +\frac{1}{k}(v_n,\,\phi_h^i)\,
    +{}\frac{1}{2} (v_n\cdot \nabla v_n,\,\phi_h^i)
    +\frac{1}{2\mathrm{Re}}(\nabla v_n,\,\nabla \phi_h^i)
    -(p_n,\,\nabla \cdot \phi_h^i) \\[5pt]
    [f_h]_i &\coloneqq  \frac{1}{k}(v_{n-1},\, \phi_h^i)
    +\frac{1}{2}(f_n,\phi_h^i)
    +\frac{1}{2}(f_{n-1},\,\phi_h^i)\\
    &\qquad
    -\frac{1}{2}(v_{n-1}\cdot\nabla v_{n-1},\,\phi_h^i)
    - \frac{1}{2\mathrm{Re}}(\nabla v_{n-1},\,\nabla \phi_h^i),
  \end{aligned}
\end{equation}
for all test functions $\phi_h^i$ and $\xi_h^i$.
The nonlinear problem is solved by Newton's method based on the initial guess $x^{(0)}=(v_{n-1},p_{n-1})$ and the iteration $l=1,2,\dots$
\begin{equation}\label{newton}
  {\cal A}'(x^{(l-1)})w^{(l)} = f_h-{\cal A}(x^{(l-1)}), \quad x^{(l)}=x^{(l-1)}+w^{(l)}
\end{equation}
where we denote by ${\cal A}'(x^{(l-1)})$ the Jacobian of ${\cal A}$ at $x^{(l-1)}$, i.e. the matrix of first partial derivatives. For the finite element discretization of the Navier-Stokes equations this is easily computed analytically, cf.~\cite[Sec. 4.4.2]{Richter2017}.

Each Newton step requires the solution of a linear system of equations, where the system matrix ${\cal A}'(x^{(l-1)})$ is sparse, but non-symmetric and not definite, due to the saddle point structure of the underlying Navier-Stokes equation. To approximate the solution with optimal robustness, we employ the generalized minimal residual method (GMRES) by Saad~\cite{Saad1996}. The convergence is thereby accelerated through a geometric multigrid solver that is used as a preconditioner with a single sweep performed for every GMRES step. We usually require about 10 linear solves for each nonlinear Newton iteration.

\subsection{The geometric multigrid method}
\label{sec:fe:mg}

\begin{algorithm}[t]
  \caption{The geometric multigrid method for the solution of the linear system $A_L \, x_L = b_L$ given on a finest level $L$.
    If all high frequency errors of $S$ are smoothed
    at a constant rate, the method achieves optimal complexity
    $O(n)$. The  multigrid solver is initiated on the finest mesh level $L$ and then used recursively:}
  \label{alg:geomg}
  \begin{algorithmic}[1]
    \Procedure{multigrid}{$l,\;A_l,\;b_l,x_l$}
    \State{$s_l \leftarrow S(A_l,\;b_l,\;x_l)$}\Comment{Smoothing}
    \State{$r_l \leftarrow b_l - A_l s_l$}\Comment{Residual}
    \State{$r_{l-1} \leftarrow \RS(l,\,r_l)$}\Comment{Restriction}
    \If{$l-1=0$}\Comment{Coarse-grid solution}
    \State{$c_0 \leftarrow A_0^{-1}r_0$}\Comment{Direct solution}
    \Else{}
    \State{$c_{l-1} \leftarrow \Call{multigrid}{l-1,\,A_{l-1},\,r_{l-1},\,0}$}
    \EndIf{}
    \State{$x'_l \leftarrow s_l + \PL(l,\,c_{l-1})$}\Comment{Prolongation}
    \State{$s'_l \leftarrow S(A_l,\;b_l,\;x'_l)$}\Comment{Smoothing}
    \State{\Return{$s'_l$}}
    \EndProcedure
  \end{algorithmic}
\end{algorithm}

Multigrid methods are a class of efficient algorithms
for the solution of linear systems that can achieve the optimal
 complexity of $O(n)$,
where $n$ is the number of degrees of freedom (in our
case proportional to the number of mesh vertices).
We employ the geometric multigrid method
as a preconditioner for the GMRES iterations.
It is summarized in Algo.~\ref{alg:geomg}.

The geometric multigrid method
is based on a hierarchical approximation
of a linear system on a sequence of finite element spaces
$V_0\subset V_1\subset \cdots \subset V_L$ defined over a
hierarchy of meshes $\Omega_0,\dots,\,\Omega_L=\Omega_h$.
Instead of solving the linear system on the finest
mesh level $L$, as one would do with traditional solvers,
geometric multigrid methods smooth high frequency
errors with a simple iterative method
and treat the remaining errors on lower levels. Usually, smoothing is
applied at the beginning (Algo.~\ref{alg:geomg}, line 2)
and at the end (Algo.~\ref{alg:geomg}, line 11) of each iteration.
In between, the
remaining residual is computed (Algo.~\ref{alg:geomg}, line 3) and restricted to the next
coarse level $L-1$ (Algo.~\ref{alg:geomg}, line 4). On this coarser level, the multigrid
iteration is then called recursively (Algo.~\ref{alg:geomg}, line 8) until the coarsest level $L-1=0$ has been reached where
a direct solver is employed (Algo.~\ref{alg:geomg}, line 6). The update from the coarse mesh
$L-1$ is then prolongated back to level $L$ (Algo.~\ref{alg:geomg}, line 10).

The mesh transfer from fine to coarse levels is accomplished with
$L^2$-projections, known as restrictions, and from coarse to fine ones
with interpolations, known as prolongations.
The smoothing that preceded the restrictions uses a
smoothing operator $S(A_l,b_l,x_l)$.
It is usually a simple iteration that yields an approximation to
the solution of the linear system, i.e.
\[
S(A_l,b_l,x_k) \approx A_l^{-1}b_l ,
\]
and that aims to quickly reduce all high frequency components of the residual $b_l-A_lx_l$.
If this happens on all levels $l$ at a constant rate, then the geometric multigrid method achieves the optimal complexity $O(n)$. In our implementation we use a simple iteration of Vanka type~\cite{Vanka1985}, which allows for easy parallelization and gives very good performance with less than 5 pre- and post-smoothing steps~\cite{FailerRichter2020,FailerRichter2021}.
The idea of the Vanka type smoother is to exactly solve small subproblems and to replace the inverse of $A_l^{-1}$ by
\begin{equation}\label{vanka0}
  {\cal V}_l(A_l) =\sum_{T\in\Omega_l}R_T^T [R_T A_lR_T^T]^{-1}R_T
\end{equation}
where $R_T$ is the restriction to those nodes that belong to one mesh element $T\in\Omega_l$ and $R_T^T$ its transpose. Considering piecewise quadratic finite elements, 9 nodes are attached to each element such that the local matrices to be inverted have the dimension $R_TA_lR_T^T \in\mathds{R}^{27\times 27}$, since each node comprises the scalar pressure and two velocity components.
\begin{equation}\label{vanka}
  S(A_l,b_l,x_k)=x_k + \omega {\cal V}_l(A_l) (b_l-A_lx_k).
\end{equation}
For a more detailed discussion of the geometric multigrid method we refer to ~\cite{Hackbusch1985,Bramble1993}
and the realization in Gascoigne 3D~\cite{Gascoigne3D} used throughout this work is described in~\cite{BeckerBraackRichter2006}.



\section{Recurrent neural networks with memory}
\label{sec:nn}

Different types of neural networks exist.
The ones originally introduced under the name are today known as
feed-forward neural networks and used for tasks such
as image recognition or data mining.
Recurrent neural networks are an extension where an activation variable
$a_n^k\in\mathbb{R}^{p_k}$ allows to propagate information in a discrete time variable indexed by $n$, such as
those occurring in time series (e.g. the stock market) or sequential data (e.g. text).
%
An extension of this model uses network nodes with memory.
These allow to more effectively model long term dependencies and also to avoid the vanishing gradient problem that otherwise occurs with recursive neural networks~\cite{Hochreiter2001}.
In our work, we use Gated Recurrent Units (GRUs)~\cite{Cho2014} that are somewhat simpler than the better known Long-Short-Term-Memory (LSTM) units~\cite{Weiss2018} but are therefore also easier to train.
At layer $k$ in the network, a GRU is given by
\begin{subequations}
  \label{eq:gru}
\begin{align}
  z_n^k &= \sigma_z \big( W_k^{(z)} \, h_n^{k-1} + U_k^{(z)} \, h_{n-1}^k + b_k^{(z)} \big)
  \\[3pt]
  r_n^k &= \sigma_r \big( W_k^{(r)} \, h_n^{k-1} + U_k^{(r)} \, h_{n-1}^k + b_k^{(r)} \big)
  \\[3pt]
  h_n^k &= z_k^t \odot h_{n-1}^k + (1 - z_n^k) \odot \sigma_h \big( W_k^{(h)} \, h_n^{k-1} + U_k^{(h)} (r_n^k \odot h_{n-1}^k) + b_k^{(h)} \big)
\end{align}
\end{subequations}
where $\odot$ denotes the element-wise product and the $W_k^{( \cdot )}$ and $b_k^{( \cdot )}$ are weight matrices and bias vectors determined by training.
The so called update gate vector $z_n^k \in \R^q$ in Eq.~\ref{eq:gru} controls to what extend $h_{n-1}^k$ is carried over to the output $h_n^k$ at the current time $n$ and the reset gate vector $r_n^k \in \R^q$ controls the contribution of $h_{n-1}^k$ to the nonlinearity of the cell.
Together, $z_n^k \in \R^q$ and $r_n^k \in \R^q$ thus control the memory of a GRU cell, i.e. to what extent the past hidden output $h_{n-1}^k$ contributes to the current one $h_n^k$.

\section{A deep neural network multigrid solver}
\label{sec:dnnmg}

In this section, we develop the deep neural network
multigrid solver (DNN-MG).
%
It uses a recurrent
neural network to predict the correction of a coarse mesh solution
that has been prolongated (or interpolated) onto one or multiple mesh levels.
Through this, we can obtain solutions that are more
accurate than those obtained with the coarse mesh only while being
computationally more efficient than performing the full (multigrid) computations
on the fine mesh level(s).

We will develop the DNN-MG solver in the context of the
Navier-Stokes equations and return to the
general formulation at the end of the section.
To simplify the exposition, we will also assume that the neural network operates on only one finer level $L+1$. However, bridging more levels is possible as we demonstrate in Sec.~\ref{sec:num:2levels}.

\begin{algorithm}[t]
  \setstretch{1.2}
  \caption{DNN-MG for the solution of the Navier-Stokes equations. Lines 6-9 (blue) provide the modifcations of the DNN-MG method compared to a classical Newton-Krylow simulation with geometric multigrid preconditioning.}
  \label{alg:dnnmg}
  \begin{algorithmic}[1]
    \For{all time steps $n$}
    \While{not converged}\Comment{Newton-GMRES method for Eq.~\ref{eq:4cranknicholson3}}
      \State{$\delta z_i \leftarrow$
        \Call{multigrid}{$L,\,A_{L}^n,\,b_{L}^n,\,\delta
          z_i$}}\Comment{Algo.~\ref{alg:geomg} as preconditioner}
      \State{$z_{i+1} \leftarrow z_i + \epsilon \, \delta z_i$}
    \EndWhile{}
    \State{\blue{$\tilde{v}_n^{\scriptscriptstyle L+1} \leftarrow \mathcal{P}(v_n^{\scriptscriptstyle L}) $}}\Comment{\blue{Prolongation on level $L+1$}}
    \State{\blue{$d_n^{\scriptscriptstyle L+1} \leftarrow \mathcal{N}(\tilde{v}_n^{\scriptscriptstyle L+1},\,\Omega_L,\,\Omega_{L+1})$}}\Comment{\blue{Prediction of velocity correction}}
    \State{\blue{$b_{n+1}^{\scriptscriptstyle L+1} \leftarrow \textbf{Rhs}(\tilde{v}_n^{\scriptscriptstyle L+1} + d_n^{\scriptscriptstyle L+1},f_n,f_{n+1})$}}\Comment{\blue{Set up rhs of Eq.~\ref{eq:4cranknicholson3} for next time step}}
    \State{\blue{$b_{n+1}^{\scriptscriptstyle L} \leftarrow \mathcal{R}(b_{n+1}^{\scriptscriptstyle L+1})$}}\Comment{\blue{Restriction of rhs to level $L$}}
    \EndFor{}
  \end{algorithmic}
\end{algorithm}



\subsection{Time stepping using DNN-MG}

\begin{figure}
  {\centering
    \makebox[\textwidth][c]{
      \begin{tikzpicture}[anchor=north west,scale=1.35,font=\small]
        \path[fill=black!5,shift={(.5,-.25)}] (0,0)
        -- node[below, midway]{$t_n$} ++(5.5,0) -- ++ (0,3.25) -- ++(-5.5,0) -- cycle;
        \path[fill=black!5,shift={(5.5,-.25)}] (0,0)
        -- node[below, midway]{$t_{n+1}$} ++(5.25,0) -- ++ (0,3.25) -- ++(-5.25,0) --
        cycle;
        \draw[black!80,dashed,shift={(5.7,-.25)}] (0,0) -- node[pos=0,anchor=north,text=black]{time step}(0,3.25);
        \draw[line width=1pt, black!50] (.5,1.8)--(10.75,1.8);
        \node[anchor= north west,font=\footnotesize] at (.5,1.825) {multigrid solution};
        \node[anchor= south west,font=\footnotesize] at (.5,1.8) {neural network};
        \draw[dotted, thick] (1,.625) -- (1.24,.625);
        \def\points{(2,0), (8,0)}
        \foreach \p in \points {
          \draw[shift={\p}] (0,0) -- (1,0) -- ++(.5,.5) -- ++(-1,0) -- (0,0);
          \draw[shift={\p},-latex] (.75,.25) -- ++(1,1);
          \draw[shift={\p},latex-] (.75,.25) -- ++(-1,1);
        }
        \def\points{(1,1), (3,1), (7,1), (9,1)}
        \foreach \p in \points {
          \draw[shift={\p}] (0,0) -- (1,0) -- ++(.5,.5) -- ++(-1,0) -- (0,0);
          \draw[shift={\p}] (0.5, 0) -- ++ (.5,.5);
          \draw[shift={\p}] (0.25,.25) -- ++ (1,0);
        }
        \def\points{(4,2), (6,2)}
        \foreach \p in \points {
          \draw[shift={\p}] (0,0) -- (1,0) -- ++(.5,.5) -- ++(-1,0) -- (0,0);
          \foreach \i in {0.25, 0.5, 0.75}{
            \draw[shift={\p}] (\i, 0) -- ++ (.5,.5);
          }
          \foreach \i in {0.125, 0.25, 0.375}{
            \draw[shift={\p}] (\i,\i) -- ++ (1,0);
          }
        }
        \draw[shift={(4,2)},latex-] (.75,.25)
        -- node[pos=.45,right=.25cm,outer sep=0,inner sep=0,fill=black!5]{
          $\mathcal{P}(v_n^{L})$
        } ++(-1,-1);
        \draw[-latex] (4.4,1.9) -- (3.5,1.9) -- ++ (1.2,1.2) -- node[above, pos=0.5]
        {temporal consistency through GRU-cell memory} (10.75, 3.1);
        \draw (.5,3.1) --node[above, pos=0.2]{$h_{t-1}$} ++(1,0) -- ++ (1.2,-1.2) --
        (3.5,1.9);
        \draw[shift={(6,2)},-latex] (.75,.25)
        --node[pos=.45,right=.25cm,outer sep=0,inner sep=0,fill=black!5]{
          $\mathcal{R}(b_{n+1}^{L+1})$
        } ++(1,-1);
        \draw[-latex,shift={(4,2)}] (1.25,.25)
        --node[above=.45cm, pos=.8,outer sep=0,inner sep=0,fill=black!5]{
          $b_n^{\scriptscriptstyle L+1}=\textbf{Rhs}
          (\tilde{v}_n^{\scriptscriptstyle L+1} + d_n^{\scriptscriptstyle L+1})$
        } ++(1,0);
        \draw[dotted, thick,shift={(9,0)}] (1,.625) -- (1.24,.625);
      \end{tikzpicture}
    }
    \caption{The neural network part of DNN-MG predicts a correction $d_n^{\scriptscriptstyle L+1}$ of the prolongated velocity $\mathcal{P}(v_n^{\scriptscriptstyle L})$ towards the unknown ground truth $\bar{v}_n^{\scriptscriptstyle L+1}$, which traditionally would be obtained using the multigrid solver for level $L+1$.
      The correction $d_n^{\scriptscriptstyle L+1}$ is incorporated into the time integration by using the improved solution $\tilde{v}_n^{\scriptscriptstyle L+1} + d_n^{\scriptscriptstyle L+1}$ to build a right hand side $b_{n+1}^{L+1}$ and restricting it to level $L$.
      The restriction $b_{n+1}^{L}$ is then used in the next time step in the multigrid solver.
       Temporal coherence of the correction (and hence the overall flow) is achieved through the memory of the GRU in the neural network.
    }
    \label{fig:rhsnnmg}
  }
\end{figure}
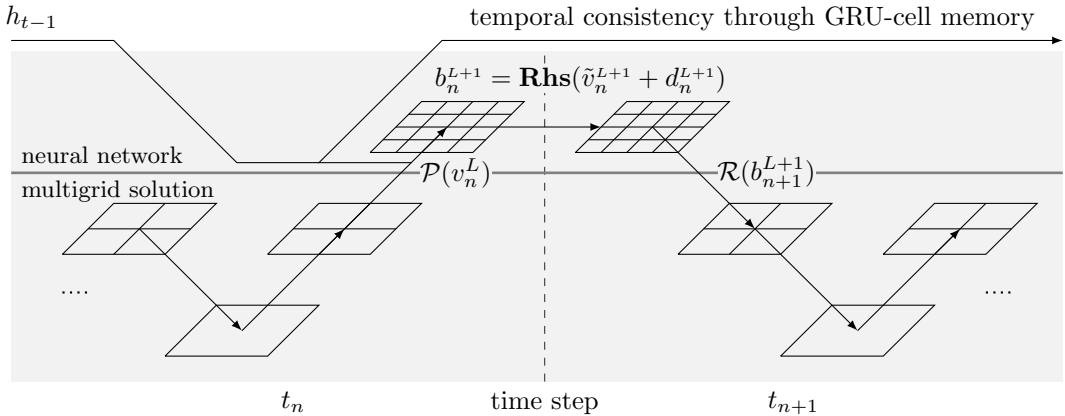

We begin by detailing one time step of the simulation of the Navier-Stokes equations
using the DNN-MG solver.
The computations are summarized in Algo.~\ref{alg:dnnmg}
and a conceptual depiction is provided in Fig.~\ref{fig:rhsnnmg}.

At the beginning of time step $n$, we first solve for the unknown velocity $v_n^{\scriptscriptstyle L}$ and pressure $p_n^{\scriptscriptstyle L}$
on the coarse level $L$ using the classical Newton-Krylow simulation
(Algo~\ref{alg:dnnmg}, lines~2-5) as described in Sec.~\ref{sec:fe:newton_krylow}.
We then interpolate (i.e. prologante) $v_n^{\scriptscriptstyle L}$ to $\tilde{v}_n^{\scriptscriptstyle L+1}$ onto the next
finer level $L+1$ (Algo~\ref{alg:dnnmg}, line~7) where a richer function space $V_{L+1}$ is available.
The neural network part of DNN-MG then predicts the velocity update $d_n^{\scriptscriptstyle L+1}$, i.e. the difference $d_{\scriptscriptstyle L+1}^n = \bar{v}_n^{\scriptscriptstyle L+1}- \tilde{v}_n^{\scriptscriptstyle L+1}$ between the prolongated $\tilde{v}_n^{\scriptscriptstyle L+1}$ and the (unknown) ground truth solution
$\bar{v}_n^{\scriptscriptstyle L+1}$ on level $L+1$ (Algo~\ref{alg:dnnmg}, line~8). 
For this,  $\tilde{v}_n^{\scriptscriptstyle L+1} \in V_{L+1}$
and information about the local mesh structure is used.
The $n^{\textrm{th}}$ time step ends by computing the right hand side $b_n^{\scriptscriptstyle L+1}$ of Eq.~\ref{eq:4cranknicholson3} on level $L+1$ using the corrected velocity $\tilde{v}_n^{\scriptscriptstyle L+1} + d_n^{\scriptscriptstyle L+1}$ (Algo~\ref{alg:dnnmg}, line~8) and restricting it to level $L$ (Algo~\ref{alg:dnnmg}, line~9).
Through this right hand side, the neural network-based correction becomes part of the multigrid computations in the next time step (Algo~\ref{alg:dnnmg}, line~5) and thus affects the overall time evolution of the flow.
This approach to build the right hand side $b_n^{\scriptscriptstyle L+1}$ on level $L+1$ and then restrict it is a key aspect of the DNN-MG solver (in fact, a restriction of  corrected velocity $\bar{v}_n^{\scriptscriptstyle L+1} + d_n^{\scriptscriptstyle L+1}$ itself would yield again the velocity $v_n^{\scriptscriptstyle L}$ that resulted from the multigrid computations only).
Pressure is handled implicitly on level $L$ in the multigrid solve and thus plays no role in the neural network DNN-MG correction.

\tikzstyle{entity} = [rectangle,
                      minimum height=1cm, anchor=south west,
                      every entity]
\tikzstyle{every entity} = [fill=black!12, align=left, inner sep=5pt]
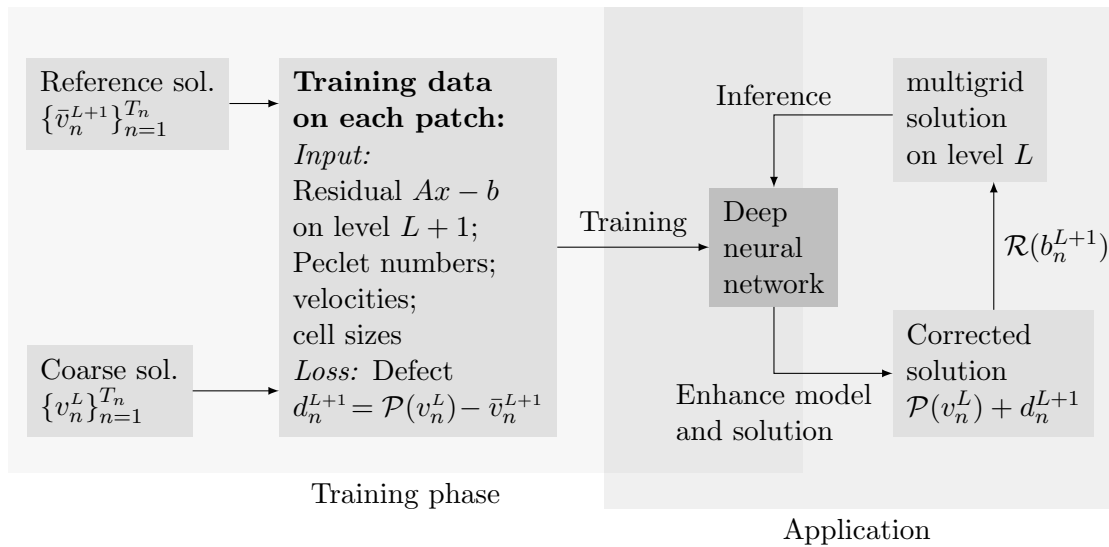
\begin{figure}[ht]
  {\centering
    \makebox[\textwidth][c]{
      \begin{tikzpicture}[node distance=2cm,scale=.95]
        \begin{scope}[transparency group]
          \begin{scope}[blend mode=multiply]
            \path[fill=black!3] (-.25,-.5)
            -- node[below, midway]{Training phase} ++(11,0) -- ++ (0,6.5) -- ++(-11,0) -- cycle;
            \path[fill=black!6] (8,-1)
            -- node[below, midway]{Application} ++(7,0) -- ++ (0,7) -- ++(-7,0) --
            cycle;
          \end{scope}
        \end{scope}
        \node[entity, minimum height=3cm](Train) at (3.5,0)
        {\textbf{Training data}\\\textbf{on each patch:}\\
          \emph{Input:} \\Residual $A x -b$ \\ on level $L+1$;\\Peclet numbers;\\velocities;\\cell sizes\\
          \emph{Loss:} Defect\\$d_n^{\scriptscriptstyle L+1} \! =\mathcal{P}({v}_n^{\scriptscriptstyle L}) \! - \bar{v}_n^{\scriptscriptstyle L+1}$};
        \node[entity] (LowFi) at (0,0) {Coarse sol.\\ $\{ v_n^{\scriptscriptstyle L} \}_{n=1}^{T_n}$ \vphantom{$V_F$}};
        \node[entity,anchor= north west] (HiFi) at (Train.north -| LowFi.west) {Reference sol.\\ $\{ \bar{v}_n^{\scriptscriptstyle L+1} \}_{n=1}^{T_n}$};
        \node[entity, right = of Train, fill=black!25] (NN) {Deep\\neural\\network};
        \node[entity] (HiFi2) at (12,0) {Corrected\\solution\\ $\mathcal{P}({v}_n^{L}) + d_n^{L+1}$};
        \node[entity, anchor=north west] (LowFi2) at
        (Train.north -| HiFi2.west) {multigrid\\solution\\on level $L$\vphantom{$V_F$}};
        \draw[-latex] (LowFi) -- node[above] {}(LowFi -| Train.west);
        \draw[-latex] (HiFi) -- node[above] {}(HiFi -| Train.west);
        \draw[-latex] (Train) -- node[above] {Training}(NN);
        \draw[-latex] (NN) |- node[below,align=left]{Enhance model\\and solution} (HiFi2);
        \draw[-latex] (HiFi2) -- node[right]{$\mathcal{R}(b_n^{L+1})$} (LowFi2.south-|HiFi2.north);
        \draw[-latex] ([yshift=.166666cm]LowFi2)
        -| node[above] {Inference} (NN);
      \end{tikzpicture}
    }
  }
  \caption{High level overview of data generation, training and
    application of DNN-MG.
    The training inputs are velocity fields on level $L$ and $L+1$, with the latter one serving as reference solution.
    From these as well as the meshes $\Omega_L$ and $\Omega_{L+1}$ on the respective levels we derive the input to the neural network that operates patch-wise (cf. Fig.~\ref{fig:network_patch}).
    At application time, we use a classical multigrid solution on level $L$ and apply the neural network part of DNN-MG to obtained a corrected solution $\mathcal{P}({v}_n^{\scriptscriptstyle L}) + d_n^{\scriptscriptstyle L+1}$ on level $L+1$. With it, a right hand side $b_n^{\scriptscriptstyle L+1}$ is computed that is then restricted to level $L$ where it is used in the multigrid solver in the next time step.
    }
    \label{fig:sketch}
\end{figure}

\subsection{The Neural Network of DNN-MG}

At the heart of the DNN-MG solver is its neural network component.
Care in its design is required to ensure that it improves the efficiency
of the runtime computations, is easy to train, and generalizes well to flow regimes
not seen during training.
These objectives will be attained through a very compact neural network with a small number
of parameters and a local, patch-based design.
An overview of the network component is provided in Fig.~\ref{fig:sketch}.

\paragraph{A patch-based neural network}
The efficiency of DNN-MG relies on the neural network evaluation
(and related auxiliary computations, e.g. of its inputs) being less computationally expensive than
using  traditional
solvers for Eq.~\ref{nonlinearshort} for level $L+1$.
When a neural network predicts the defect $d_n^{L+1}$ for the entire domain on level $L+1$,
the size of the networks in- and outputs is of order $O(m_{L+1})$
where $m_{L+1}$ is the number of degrees of freedom on the level.
The cost of evaluating the network is then at least
$O(m_{L+1}^2)$ and it would hence
be more expensive than a direct multigrid solution.
The neural network would additionally have to have a large
number of parameters to model the solution's behavior on the entire simulation domain,
necessitating substantial training time and data.
The domain would also be ``baked'' into the network, hampering generalization.

\begin{figure}
  \centering
  \begin{tikzpicture}[font=\scriptsize]
    \path[draw=black!15, fill=black!5,rounded corners=6pt] (0,0) -- (6,0) -- ++(2,2) --
    ++(-6,0) -- cycle;
    \node[anchor=north west] at (5.9,1.9){domain $\Omega$};
    \foreach \i in {0,.75,1.5}{
      \draw[shift={(1.625,0.125)},dotted] (\i, 0) -- ++ (1.5,1.5);
      \draw[shift={(1.25,.5)},dotted] (.5*\i, .5*\i) -- ++ (3,0);
      \draw[shift={(2,.5)}] (\i, 0) -- ++ (.75,.75);
      \draw[shift={(2,.5)}] (.5*\i, .5*\i) -- ++ (1.5,0);
    }
    \foreach \i in {0,1.5}{
      \draw[shift={(2,.5)}] (\i, 0) -- ++ (.75,.75);
      \draw[shift={(2,.5)}] (.5*\i, .5*\i) -- ++ (1.5,0);
    }
    \node[fill=black!20,align=left,font=\scriptsize,inner sep=2pt] (NN) at (10, 0.75) {GRU-based Neural\\Network};
    \node[align=left,anchor=north west,fill=black!5,inner
    sep=0pt,font={\bfseries\scriptsize},trapezium, trapezium left angle=45, trapezium right angle=135]
    at (4.6, 1.2){Patch\\\hspace{-.375cm}from mesh};
    \draw[-latex,shorten <=.66cm] (3.125, 0.875)
    to[out=45,in=135] node[fill=white,align=left,midway, above=-0.33cm]{Get input\\for DNN\\from patch}
    (NN.north);
    \draw[latex-,shorten <=.66cm] (3.125,0.875)
    to[out=-135,in=-135] node[pos=.66,fill=white,align=left]{Add defect\\back to\\solution}
    (NN.south);
  \end{tikzpicture}
  \caption{DNN-MG uses a local approach where the neural network operates patch-wise on small neighborhoods of the simulation domain and provides a velocity correction independently for each patch.}
    \label{fig:network_patch}
\end{figure}
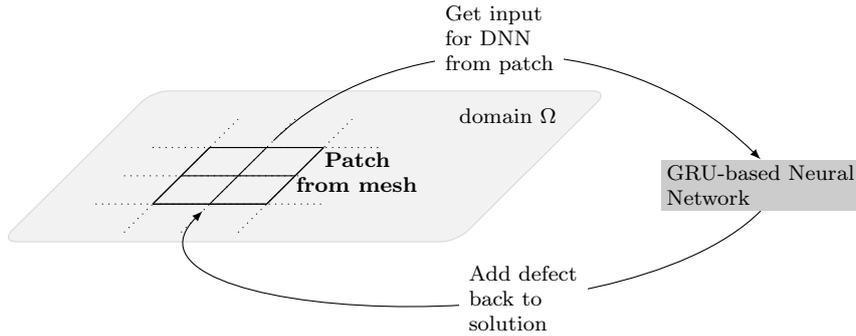

To avoid these issues, DNN-MG uses a neural network that
operates patch-wise over small neighborhoods
of the mesh. In its most local variant each patch consists of the unknowns that belong to one single mesh element only, cf. Fig.~\ref{fig:network_patch}; a more global approach is easily realized by considering larger patches consisting of multiple adjacent elements.
Our network takes as input thus only information from one
patch and predicts the velocity update $d_n^{\scriptscriptstyle L+1}$
only for the mesh nodes in the patch, with the prediction done independently
for all patches to cover the entire
simulation domain
(duplicate predictions for adjacent patches are averaged).
The locality is one key to the compactness of the neural network and hence also to its efficient evaluation.
Furthermore, since a single neural network is used for all patches in the domain,
training with a small number of flows exposes the network
to a large number of different flow behaviors.
This facilitates the network's ability to generalize to flows not seen during training and reduces training
time and data.

\paragraph{Neural network inputs}
The efficacy of the neural network prediction relies critically on suitable network inputs that
provide rich information about the coarse flow behavior on level $L$ as well as the local patch geometry
(similar to how the geometry of mesh cells play an important role in classical
finite element simulations).
In particular, with carefully selected inputs one can potentially reduce the number of parameters in the
neural network, which reduces the evaluation time at runtime and the amount of data and time
required for training.

When the most localized patches directly supported by the mesh are used (as we do in most of our numerical experiments) each patch consists of exactly one mesh element on level $L$ with $N_L=4$ nodes and thus with $N_{L+1}=9$ nodes on level $L+1$.
Our inputs to the neural network are then:
\begin{itemize}
  \item nonlinear residuals $r_n^{\scriptscriptstyle L+1}=f_{\scriptscriptstyle L+1}-{\cal A}_{\scriptscriptstyle L+1}({\cal P}(x_n^{\scriptscriptstyle L}))\in \R^{3 \, N_{L+1}}$ of the prolongated coarse mesh solution for Eq.~\ref{nonlinearshort};
  \item the velocities $v_n^{L} \in \R^{2N_L}$ on mesh level $L$;
  \item P{\'e}clet numbers $\mathrm{Pe}_{L} \in \R^{N_L}$;
  \item geometry of the cells, in particular the edge lengths $h^c\in \R^4$,
  the aspect ratios $a^c\in \mathds{R}^4$ (of two neighboring and two opposite edges each), and the angles between the edges $\alpha^c\in\mathds{R}^4$ .
\end{itemize}
We hence have in total $51$ inputs to the neural network.

The most important input is the nonlinear residual
$r_n^{\scriptscriptstyle L+1}=f_{\scriptscriptstyle L+1}-{\cal A}_{\scriptscriptstyle L+1}({\cal P}(x_n^{\scriptscriptstyle L}))$, cf. Eq.~\ref{nonlinearshort}, that provides a measure of the local error (or defect) at every mesh node on level $L+1$ and thereby incorporates information about the partial differential equation.
We also use the P{\'e}clet number $\mathrm{Pe}_{L} = q \, \tilde{v}_{L}^n / \nu$ where
$\nu$ is the fluid's viscosity and $q$ a characteristic length scale, which is the patch size in our case.
The P{\'e}clet number describes the ratio between transport and diffusion on each node on level $L$ and it hence measures which part of the Navier-Stokes equations dominates the local behavior of the flow.

\paragraph{Neural network architecture}
To predict a defect $d_n^{\scriptscriptstyle L+1}$ that is consistent
with the fluid flow, we use recurrent neural
networks with memory.
Through this, the past simulation states $\,\dots,\,v_{n-2},\,v_{n-1}$ affect
the predicted update $d_n$ at time step $n$.

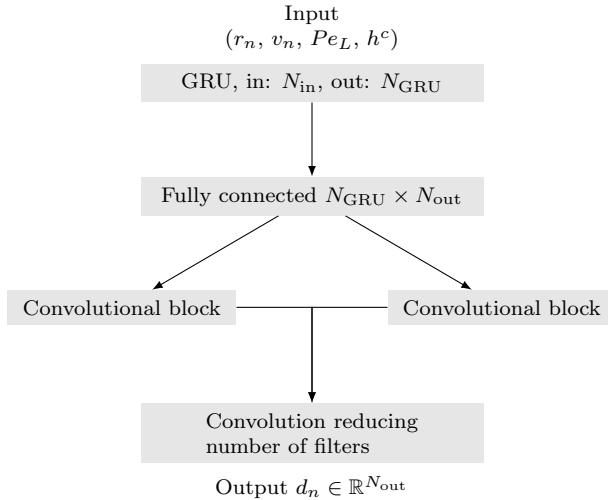
\begin{figure}[t]
  \centering
  \begin{tikzpicture}[transform shape]
    \tikzstyle{every node}=[font=\scriptsize,minimum width=4.5cm,fill=black!10, anchor=north west]
    \node (gru) at (0,0){GRU, in: $N_{\textrm{in}}$, out: $N_{\textrm{GRU}}$};
    \node[fill=white,above=1pt of gru,align=center] (in){Input\\$(r_n,\,v_n,\,Pe_L,\,h^c)$};
    \node[minimum width=4.5cm] (fc) at (0,-1.5){Fully connected
      $N_{\textrm{GRU}}\times N_{\textrm{out}}$};
    \draw[-latex] (gru) -- (fc);
    \node[align=left,minimum width=3cm] (convv) at (-1.75,-3){Convolutional block};
    \node[align=left,minimum width=3cm] (convl) at (3.25, -3){Convolutional block};
    \draw[-latex] (fc) -- (convv);
    \draw[-latex] (fc) -- (convl);
    \node[align=left] (convlast) at (0,-4.5) {Convolution reducing\\number
      of filters};
    \draw[-latex] (convv) -| (convlast);
    \draw[-latex] (convl) -| (convlast);
    \node[fill=white,below=1pt of convlast,align=center] (out){Output $d_n\in\R^{N_{\textrm{out}}}$};
  \end{tikzpicture}
  \caption{The neural network architecture that was used.}
  \label{fig:nn_arch}
\end{figure}

The specific network architecture we use for DNN-MG is shown in
Fig.~\ref{fig:nn_arch}.
At its heart is a GRU unit with memory, cf. Sec~\ref{sec:nn}.
It could in principle be replaced by another cell with
memory, such as an LSTM unit or a TCN, but since the GRU unit performed
well in our experiments and is, through its simpler design, easier to train we used it
in the present work.
%
The two convolutional blocks in the network are designed
to learn local dependencies, with one
learning the interactions of the two velocity components at each
vertex and the other the spatial dependence of
each component across a patch.
Each of these blocks consist of two layers, where the second one is
just the transposed version of the first layer so that input and output sizes are equal.
%
The final convolutional layer reduce the dimensionality of the data from the
multiple filters that form the preceding convolutional layers to the output $d_n$
of the network.

\paragraph{Training of DNN-MG}

The training of DNN-MG is based on simulations of the Navier-Stokes equations for which a
multigrid representation of the velocity $v_n^l$ with two levels $L$ and $L+1$ is available.
The velocity $v_n^{L+1}$ thereby serves as ground truth $\bar{v}_n^{\scriptscriptstyle L+1}$.
The training then finds network weight (in the GRU unit, the dense and the convolutional layers) such that
the norm $\big\Vert (\tilde{v}_n^{\scriptscriptstyle L+1} + d_n^{\scriptscriptstyle L+1}) - \bar{v}_n^{\scriptscriptstyle L +1 } \big\Vert$ of the difference
between the predicted velocity $\tilde{v}_n^{\scriptscriptstyle L+1} + d_n^{\scriptscriptstyle L+1}$ (i.e. the velocity after correction, Algo.~\ref{alg:dnnmg}, line 9) and the ground truth $\bar{v}_n^{\scriptscriptstyle L}$ is minimized over the course of a simulation.

\usetikzlibrary{decorations.pathreplacing} \usetikzlibrary{fadings}

\subsection{A general formulation of DNN-MG}
\label{sec:dnnmg:general}

%
%
%
%
%
%
After the detailed description of DNN-MG for the solution of the Navier-Stokes equations we will in the following show
how it can be applied to a general PDE in variational form
$u\in \mathcal{V}\colon\; A(u)(\Phi) = F(\Phi), \ \forall \Phi \in \mathcal{V}$,
where $A(\cdot)(\cdot)$ is a semi-linear form (linear
in the second argument) and $\mathcal{V}$ is a Banach space of test functions for the problem. Given a suitable Galerkin formulation in the discrete subspace $V_h\subset \mathcal{V}$ the abstract Newton iteration for the solution of the problem ${\cal A}_h(x_h)=f_h$ can be formulated exactly as for the Navier-Stokes equations, compare Eq.~\ref{nonlinearshort} and Eq.~\ref{newton}. The Newton-Krylov approach based on the preconditioned GMRES iteration and a geometric multigrid solver as preconditioner is highly robust and it can be applied to a variety of PDE models (as is done in Gascoigne~\cite{Gascoigne3D}).
The DNN-MG solver in Algorithm~\ref{alg:dnnmg} can hence in principle be applied to various PDEs.


\subsection{Algorithmic Complexity of DNN-MG}

The Newton-Krylov geometric multigrid method can already achieve optimal complexity, which raises the question what advantage DNN-MG has to offer. Before addressing the question from a practical point of view in the next section, we sketch in the following an answer from a theoretical point of view.

The Newton-Krylov geometric multigrid framework achieves linear complexity $\mathcal{O}(N)$ in the number $N$ of degrees of freedom, which roughly quadruples with each global mesh refinement, i.e. $N^{L+1}\approx 4N^L$. The constant hidden in $\mathcal{O}(N)$, however, can be very significant, since on average 5 Newton steps are required in each time step and within each Newton step one has to compute on average 10 GMRES steps with one sweep of the geometric multigrid solver as preconditioning.
Furthermore, in the geometric multigrid typically 5 pre-smoothing and 5 post-smoothing steps have to be determined.
Hence, a total of about 500 Vanka smoothing steps
must be performed on each of the mesh layers $L,L-1,\dots,0$. One Vanka step thereby requires the inversion of about $\mathcal{O}(N^l/4)$ small matrices of size $27\times 27$, one on each element of the mesh, cf. Eq.~\ref{vanka0}, resulting in approximately $27^3\approx 20\,000$ operations. Since the complexity of all mesh levels sums up to about $N^L + N^{L-1}+\dots N_0 \approx N^L(1+4^{-1}+\dots + 4^{-L})\approx \frac{4}{3}N^L$ we can estimate the effort for the complete solution process on level $L$ as $5\cdot 10\cdot  (5+5)\cdot 27^3\cdot \frac{4}{3}\approx 10^7 N^L$. We thereby only counted the effort for smoothing and neglected all further matrix, vector and scalar products.
Solving the problem on an additional mesh level $L+1$ would require
about four times this effort, i.e. $\approx 4\cdot 10^7 N^L$ operations.

On the other hand, the DNN-MG method only requires the prolongation of the solution to the
next finer mesh and there, as main effort, the evaluation of the neural network on each
patch. If we again consider patches of minimal size (one patch is one element of mesh layer
$L+1$) about $\mathcal{O}(N^{L+1}/4)=\mathcal{O}(N^L)$ patches must be processed. The
effort for the evaluation of the network can be estimated by the number of trainable
parameters with
$N_c$ inputs. The DNN-MG approach asks for only one such evalution of the network in each
time step. The specific network model used for the numerical test cases comprises 8634
trainable parameters. Hence, the effort for correcting the level $L$ Newton-Krylov
solution by the neural network on level $L+1$ can be estimated by $8634 N^L$, which is negligible in comparison to the effort of solving on level $L+1$ ($\approx 40^7 N^L$).

\section{Numerical examples}
\label{sec:num}

\begin{figure}[!htb]
  \centering
  \begin{tikzpicture}[scale=5.5]
    \draw [fill=gray!20] (0,0) rectangle (2.2,0.41);
    \path[draw=black] (0,0) -- node[midway, above]{$\Gamma_{\textrm{wall}}$} ++(2.2,0) --
    node[midway, left]{$\Gamma_{\textrm{out}}$} ++(0,0.41) -- node[midway,
    below]{$\Gamma_{\textrm{wall}}$} ++(-2.2,0) -- node[midway, right]{$\Gamma_{\textrm{in}}$}cycle;
    \draw[fill=white] (0.2,0.2) ellipse (0.04 and 0.06);
    \node at (1.5, 0.2){$\mathrm{Re}_{\text{test}}= 200$};
    \draw[decorate,decoration={brace,raise=1pt,amplitude=9pt,mirror}] (0,0) --
    node[below=9pt]{$2{.}2$} (2.2,0);
    \draw[decorate,decoration={brace,raise=1pt,amplitude=9pt}] (0,0) --
    node[above left=12pt and 12pt,rotate=90]{$0{.}41$} (0,0.41);
    \draw[decorate,decoration={brace,raise=1pt,amplitude=7pt,mirror}] (0.24,0.14) --
    node[right=7pt]{$0{.}12$} (0.24,0.26);
    \draw[decorate,decoration={brace,raise=1pt,amplitude=7pt,mirror}] (0.16,0.14) --
    node[below=7pt]{$0{.}08$} (0.24,0.14);
    \draw[-latex] (0.2, 0.3) node[above]{$(0{.}2,\,0{.}2)$} --(0.2,0.2);
  \end{tikzpicture}
\caption{Geometry of the test scenario with a parabolic inflow profile $\Gamma_{\textrm{in}}$,
  do-nothing boundary conditions at the outflow boundary $\Gamma_{\textrm{out}}$ and no-slip conditions on the walls
  $\Gamma_{\textrm{wall}}$. The center of the obstacle is at $(0.2,\,0.2)$.}
  \label{fig:geom}
\end{figure}
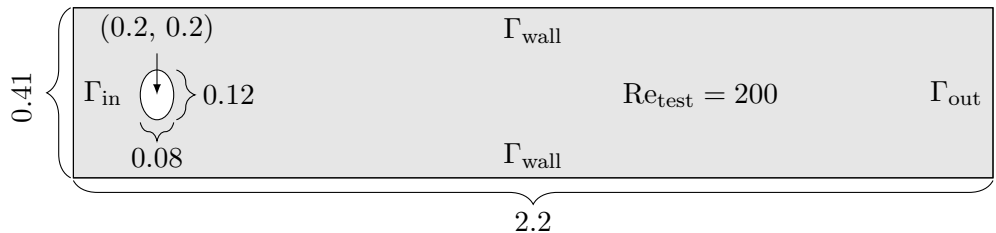

In this section, we present numerical results obtained with the deep neural network multigrid solver (DNN-MG) for the Navier-Stokes equations.
We consider a classical benchmark problem describing the laminar flow around a cylinder, cf.~\cite{SchaeferTurek1996}. In contrast to the original benchmark configuration, we parametrize the geometry with an ellipsoid with varying aspect ratio as obstacle. This allows us to easily generate different but similar training data and also provides a basic test for the generalization ability of our technique.
We also report on the runtime performance of DNN-MG and study its ability to generalize to flow regimes that differ from the training data.
We will work with DNN-MG with the classical simulation operating up to level $L$ and the neural network  on level $L+1$ except for Sec.~\ref{sec:num:2levels} where we will consider prediction across two levels and with larger patches.

\subsection{Setup}
\label{sec:num:setup}

We consider a variation of the
well-established ``laminar flow around a cylinder'' introduced by Sch\"afer and
Turek~\cite{SchaeferTurek1996}, in particular the unsteady 2D-2
testcase where the laminar flow around
a circular obstacle at Reynolds number $\mathrm{Re}=100$ leads to a stable
periodic flow pattern.
The geometry of the flow domain is shown in Fig~\ref{fig:geom}.
The flow is driven by a Dirichlet
profile $v=v^D$ at the left boundary $\Gamma_{\mathrm{in}}$ given by
\begin{equation}
 \begin{aligned}
v^D(x,y,t) &= v_{avg}\frac{6y(H-y)}{H^2}\omega(t) \ \text{  on  } \ \Gamma_{in}\coloneqq
0\times [0,\,H],
\\[4pt]
\; \omega(t) &= \begin{cases} \frac{1}{2}- \frac{1}{2}\cos( 2\pi t), & t\le \frac{1}{2} \\ 1 & t>\frac{1}{2},
\end{cases}
  \end{aligned}
\end{equation}
where $H=0.41$ is the height of the
channel and the function $\omega(t)$ regularizes the startup phase of the
flow. On the wall boundary $\Gamma_{wall}$ and on the
obstacle we prescribe no-slip boundary conditions $v=0$ and on the outflow
boundary $\Gamma_{\mathrm{out}}$ we use a do-nothing outflow condition~\cite{HeywoodRannacherTurek1992}.
The average flow rate is $v_{avg}=\frac{4}{3}$ which, through the different obstacles used for training and testing, results in the Reynolds
numbers $\mathrm{Re}_\text{train}=166$ and $\mathrm{Re}_\text{test}=200$, respectively.

In all numerical examples the temporal step size is
$k=0{.}01$. In space, we consider the sequence of meshes in
Table~\ref{tab:mesh}. Finer levels result from a uniform refinement
of the ``coarse'' mesh conforming to the obstacle (see~\cite{Gascoigne3D} for details on conforming boundary handling).



\begin{table}[t]
  \begin{center}
    \begin{tabular}{lrrl}
      \toprule
      & unknowns & \multicolumn{2}{r}{multigrid levels} \\
      \midrule
      coarse        & 2124  & 3 & ($L$)\\
      fine          & 8088  & 4 & ($L+1$)\\
      reference     & 31536 & 5 & ($L+2$)\\
      \bottomrule
    \end{tabular}
  \end{center}
  \caption{Spatial meshes with increasing number of multigrid levels
    and unknowns.}
  \label{tab:mesh}
\end{table}

\subsection{Neural network parameters}

As discussed in Sec.~\ref{sec:dnnmg}, DNN-MG uses a local neural network that operates over patches of the mesh domain.
For our numerical experiments, we defined a patch to be the refinement of a coarse mesh
cell on level $L$, i.\,e.\ we used the most localized patches that are naturally supported by
the mesh structure.
This resulted in a network (cf.\ Figure~\ref{fig:nn_arch}) with only $8634$ trainable parameters.
With none of the layers having a bias and
$N_{\textrm{in}}=51$ (the input length),
$N_{\textrm{GRU}}=32$ (the size of the GRU state),
$N_{\textrm{out}}=18$ (the output length),
there are $3(32\cdot 32+32\cdot 51)=7968$ parameters in the GRU
and $32\cdot 18= 576$ in the fully connected layer.
Each of the convolutional blocks that learns vertex interactions has, furthermore, 36 parameters and the last
convolutional layer has 18 parameters.

\blue{
Note that the length of the feature vector (input length)
only depends on the number of elements that make up a patch but not on
the number of elements or patches in the mesh. This is due to our
local patch-based design and in particular, the network can be used independently of the number of elements, refinement
level of the mesh, the domain $\Omega$ and the overall flow situation.
However, changing these parameters too much could affect the performance of the network.
This is closely linked to the generalizability, which we demonstrate in
Section~\ref{sec:num:generalizability}.
}


Each convolutional block learns 4 filters, in the first layer the number of filters
is expanded to 4 and in the following transposed layer it is reduced to 1.
In the last layer the
outputs of the convolutional blocks are concatenated along the filter dimension,
which then has to be reduced to a length of 1.
This is achieved by a a convolution with length 1 (summing along the
filter dimension and weighting the output),
resulting in 18 trainable parameters.

The network was implemented using PyTorch~\cite{PyTorch}.

\subsection{Neural network training}
\label{sec:num:nn:training}


As training data for DNN-MG we used three $(L+1)$-level multi-grid simulations of the flow around a cylinder, each with a slightly different elliptical obstacle with height $0{.}1$.
The small number of simulations was sufficient since, by the
patch-wise application of the network, a single simulation
provides $N_c \times N_T$ training items, where $N_c$ is the number of patches
and $N_T$ is the length of the time series.
By the periodicity of the flow it also sufficed to
consider a small number of full periods.
We hence used $t \in [1,\,6]$ resulting in $N_T=500$ (we used $t > 1$ to avoid the start-up phase of the simulation where the quasi-periodic flow forms).

As loss function $\mathcal{L}$ we employed a simple
$l^2$-loss
\[
  \mathcal{L} =
  \sum_{n=1}^{N_T} \frac{1}{N_c} \sum_{c=1}^{N_c} l(\tilde{v}_{n,c}^{\scriptscriptstyle L+1} + d_{n,c}^{\scriptscriptstyle L+1} ,\,\bar {v}_{n,c}^{\scriptscriptstyle L+1})
  = \sum_{i=1}^{N_T} \frac{1}{N_c} \sum_{c=1}^{N_c} \norm{ \big( \tilde{v}_{n,c}^{\scriptscriptstyle L+1} + d_{n,c}^{\scriptscriptstyle L+1} \big) - \bar{v}_{n,c}^{\scriptscriptstyle L+1} }_2 .
\]
Tikhonov-regularization was used with a scaling factor $\alpha=10^{-4}$.
Training was performed with the ADAM optimizer with a limit of 1000 epochs.
At the end, the model with the lowest
validation loss was saved.
This resulted in a
training time of 4h 23m on an Intel Xeon Gold 6254.

\subsection{Flow prediction}
\label{sec:flowprediction}

\begin{figure}[!htb]
  \centering
      \includegraphics[width=\linewidth]{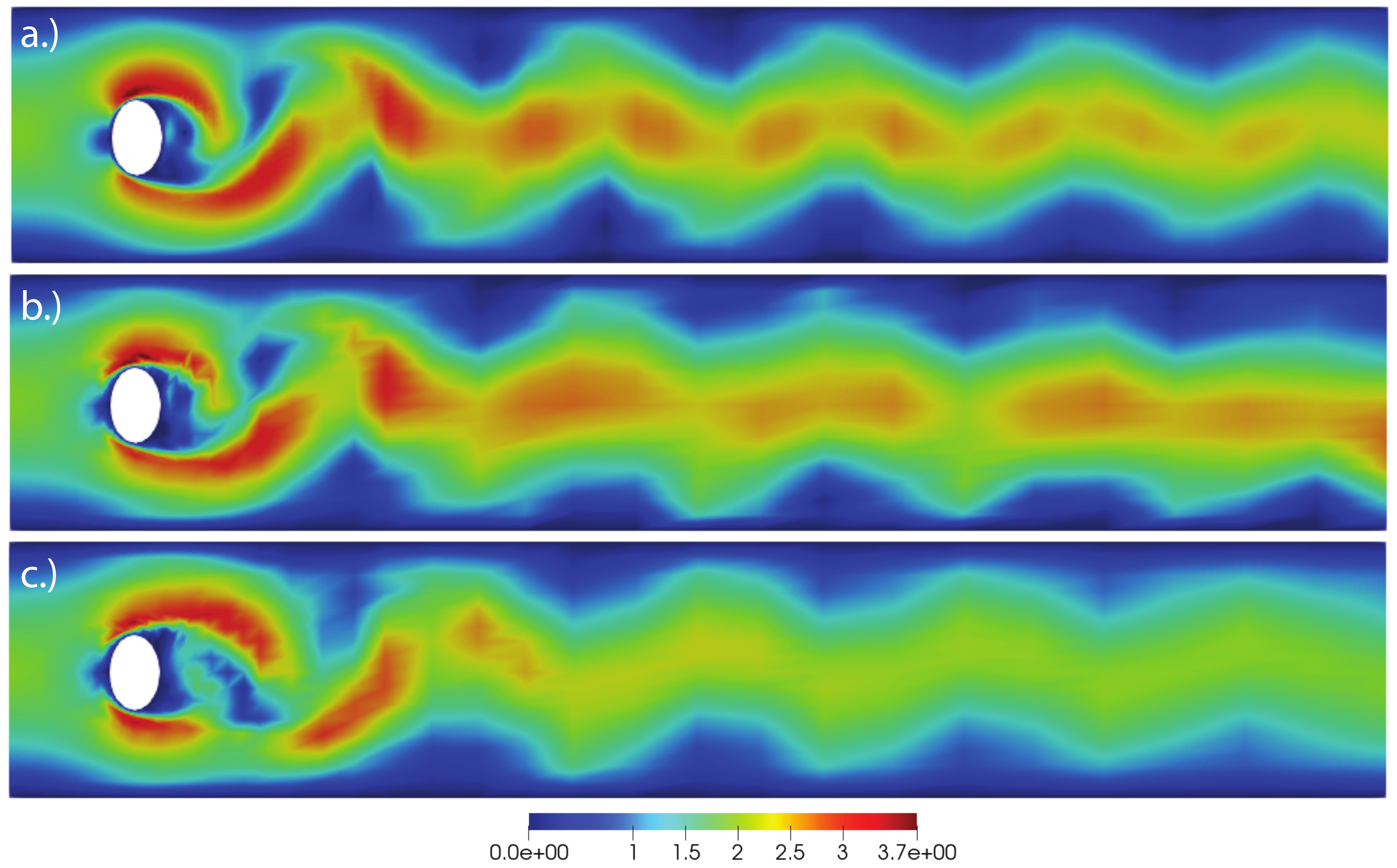}
      \caption{Magnitude of the velocity field for the channel for flow with one obstacle at time $t=9$ for a.) a multigrid solution with $L+1$ levels, b.) DNN-MG, c.) a multigrid solution with $L$ levels.}
      \label{fig:velocityfields}
\end{figure}

For testing we used the flow around an elliptic obstacle
with an increased height of $0.12$ compared to the training data (see Fig.~\ref{fig:geom}).
Fig.~\ref{fig:velocityfields} shows the velocity magnitude at time
$t=9$ for the flow obtained using DNN-MG as well as classical multigrid solutions on level $L$ and $L+1$.
It can be observed that DNN-MG is indeed able
to predict high frequency fluctuations that are not visible in the coarse
mesh solution. In particular in the vicinity of the obstacle the quality of the
solution is strongly enhanced with distinct features in the wake being apparent
in the DNN-MG simulation.

\begin{figure}[!htb]
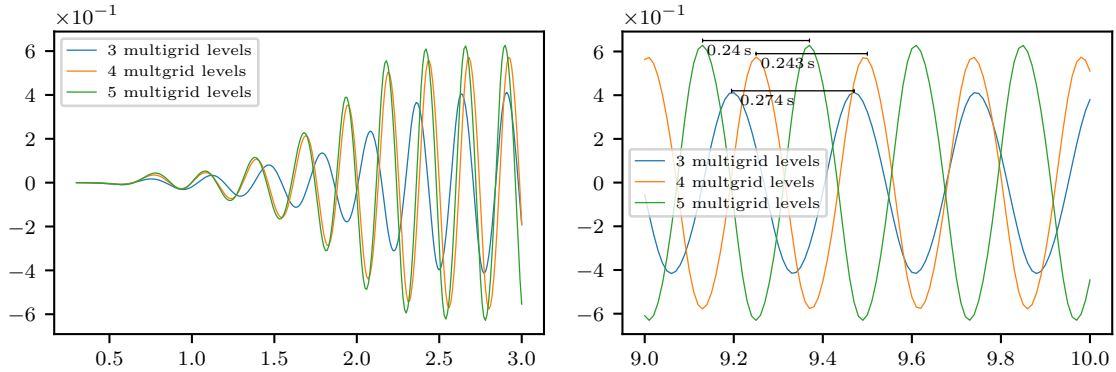

  \scalebox{.97}{\input{plots/liftfcts1_3.pgf}}
  \scalebox{.97}{\input{plots/liftfcts9_10.pgf}}
  \caption{The lift functional on the sequence of spatially refined
    finite element meshes from Table~\ref{tab:mesh} reveals a frequency shift which is depending on the spatial finite element discretization. We show the
    startup phase $[0,3]$ and the interval $[9,10]$, where the flow is
    fully developed.}
  \label{fig:liftfreq}
\end{figure}

\begin{figure}[!htb]
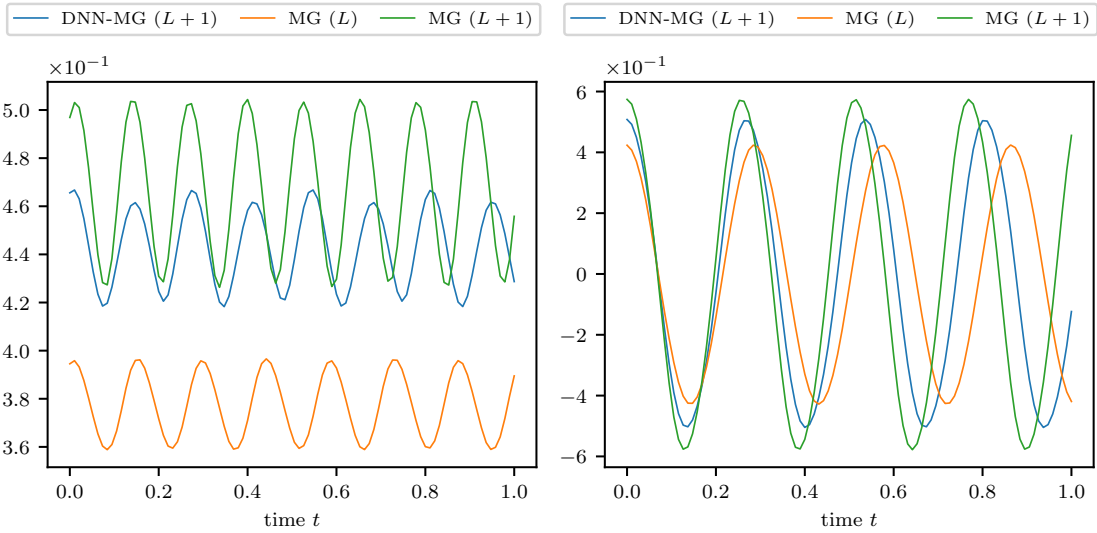

  \centering
  {\centering
  \resizebox{0.49\textwidth}{!}{\input{plots/dragfcts_convminimal.pgf}}
  \resizebox{0.49\textwidth}{!}{\input{plots/liftfcts_convminimal.pgf}}}
  \caption{Drag (left) and lift (right) functionals for the channel with one obstacle for the coarse
    mesh solution, the level $L+1$ solution and DNN-MG($L+1$).}
  \label{dl2D_min}
\end{figure}

For a quantitative analysis of the accuracy of DNN-MG,
 we compare the drag and lift forces on the
obstacle,
\[
J_d(v,p) = \int_\Gamma \Big(\frac{1}{\mathrm{Re}} \nabla v - pI\Big)\vec n \cdot \vec e_1 \,\text{d}s,\quad
J_l(v,p) = \int_\Gamma \Big(\frac{1}{\mathrm{Re}} \nabla v - pI\Big)\vec n\cdot
\vec e_2\,\text{d}s,
\]
respectively, where $\vec e_1=(1,0)^T$ and $\vec e_2=(0,1)^T$.
The results in Fig.~\ref{dl2D_min} and Table~\ref{tab:draglift} demonstrate that DNN-MG
is able to substantially correct both the drag and lift functionals.
Maximum and
minimum values but also the frequency of oscillation of the augmented
simulation are significantly closer to the fine mesh
one that serves as reference.

\begin{figure}[t]
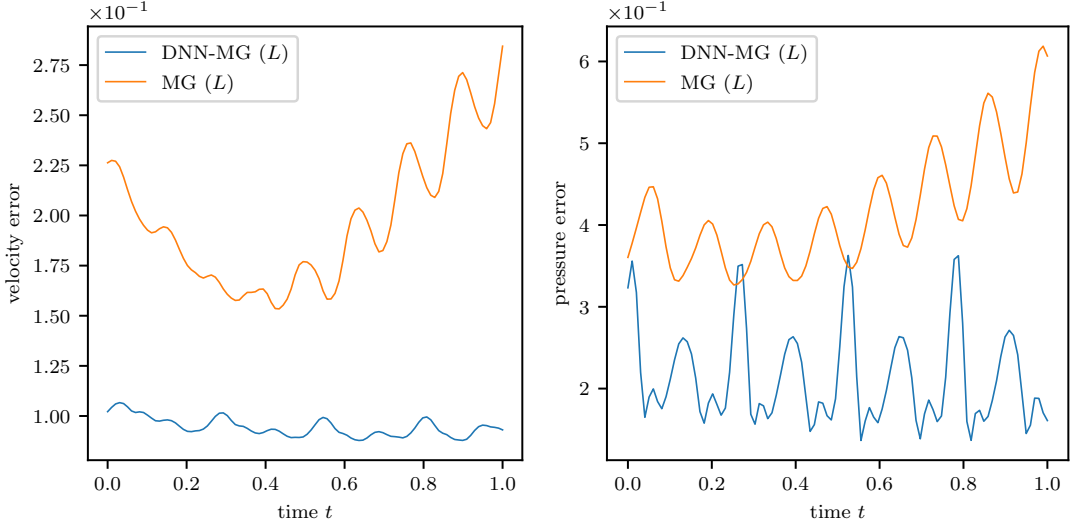

  \centering
  \resizebox{0.48\textwidth}{!}{\input{plots/VResults2d2_convminimal.pgf}}
  \resizebox{0.48\textwidth}{!}{\input{plots/PResults2d2_convminimal.pgf}}
  \caption{
    Relative errors for coarse solution with and without ANN correction
    compared to the reference solution on level $L+1$.}
  \label{fig:velpres}
\end{figure}

Fig.~\ref{fig:velpres}, finally, shows the relative velocity and pressure errors
\[
E_n^{\textrm{rel}}(v) = \frac{\big\Vert (\tilde{v}_n^{\scriptscriptstyle L+1} + d_n^{\scriptscriptstyle L+1} ) - v_n^{\scriptscriptstyle L+1} \big\Vert_2}{\big\Vert v_n^{\scriptscriptstyle L+1}  \Vert_2},\qquad
E_n^{\textrm{rel}}(p) = \frac{\big\Vert p_g-p_f \big\Vert_2}{\big\Vert p_n^{\scriptscriptstyle L+1} \big\Vert_2},
\]
where $\Vert \cdot \Vert_2$ denotes the Euclidean norm of the discrete
solution.
DNN-MG is able to robustly reduce the velocity error from
$20\%-27\%$ to about $10\%$. While the relative pressure error is also reduced by a
factor of 2 on average, some oscillatory peaks remain. These
stem from the shift in periodicity as discussed in the following remark.

\begin{remark}
Fig.~\ref{fig:liftfreq} shows the lift functional for classical multigrid simulations
with three different resolutions both in the startup phase $t\in [0,3]$
and for $t \in [9,10]$ where the flow profile is fully developed.
It can be seen that on coarser meshes it
takes longer for the periodic flow profile to develop
(Fig.~\ref{fig:liftfreq}, left) and the period of oscillations is getting shorter with
increasing mesh resolution (Fig.~\ref{fig:liftfreq}, right).
This shift
in frequency and phase as a function of mesh resolution prevents
a direct comparison of results obtained on different levels in a multigrid-type algorithm such as DNN-MG.
In~\cite{MargenbergRichter2020}, a shifted variant of the Crank-Nicolson scheme is discussed to
remedy the problem.
Since this cannot be used directly for DNN-MG,
we instead compare the functional values in shifted intervals
$[t_*,t_*+1]$, where $t_*$ is chosen as the first peak in the lift
functional after time $t=9s$, identified separately for the coarse
mesh solution, DNN-MG and the fine mesh one on level $L+1$.
The results reported in Fig.~\ref{dl2D_min}, Fig.~\ref{fig:velpres},
Fig.~\ref{dl2D_min2obs} and Table~\ref{tab:draglift}, Table~\ref{tab:draglift:double}
are those for the temporally adjusted flow in $[t_*,t_*+1]$.
\end{remark}

\begin{figure}[!htb]
  \centering
  \resizebox{0.48\textwidth}{!}{\input{jmp2-2/dragfcts_ctest.pgf}}
  \resizebox{0.48\textwidth}{!}{\input{jmp2-2/liftfcts_ctest.pgf}}
  \caption{
    Drag (left) and lift (right) functionals for the channel with one obstacle for the coarse
    mesh solution, the level $L+2$ solution and DNN-MG ($L+2$).}
  \label{fig:dragliftjmp2}
\end{figure}

\begin{table}[t]
  \centering
  \resizebox{\textwidth}{!}{
  \begin{tabular}{l|llll|lllll}
    \toprule
    &\multicolumn{4}{c|}{drag}&\multicolumn{4}{c}{lift}\\
    type (level) & min    & max    & mean & ampl.& min    & max    & mean & ampl.\\
    \midrule
    MG ($L+2$) & $0{.}4566$ & $0{.}5305$ & $0{.}4935$& $0{.}0739$& $-0{.}5680$& $0{.}5648$& $-0{.}0016$& $1{.}1328$\\
    DNN-MG ($L+2$) & $0{.}4521$ & $0{.}5112$ & $0{.}4842$& $0{.}05908$& $-0{.}5708$& $0{.}5697$& $-0.0085$& $1{.}1405$\\
    MG ($L+1$) & $0{.}4263$ & $0{.}5045$ & $0{.}4654$& $0{.}07816$& $-0{.}5779$& $0{.}5742$& $-0{.}00185$& $1{.}1521$\\
    DNN-MG ($L+1$) & $0{.}4179$ & $0{.}46674$ & $0{.}4424$& $0{.}04868$& $-0{.}5047$& $0{.}5081$& $ \phantom{-}0{.}00187$& $1{.}0129$\\
    MG ($L$) & $0{.}3699$ & $0{.}4075$ & $0{.}3887$& $0{.}03766$& $-0{.}4162$& $0{.}4130$& $-0{.}00163$& $0{.}8291$\\
    \bottomrule
  \end{tabular}}
  \caption{Functional outputs for the drag and the lift functional on
    the coarse mesh, the fine mesh and the coarse  mesh with ANN correction.}
  \label{tab:draglift}
\end{table}

\subsection{Timings}

\begin{figure}[t]
  \resizebox{\textwidth}{!}{\input{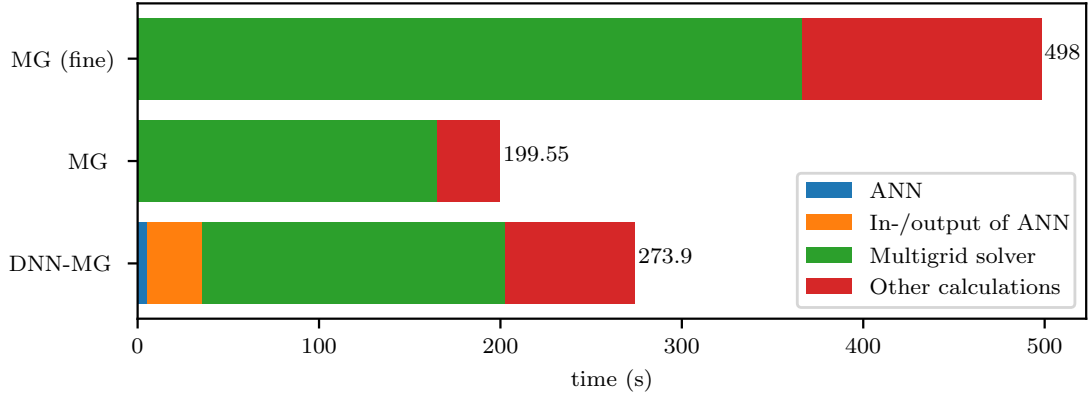}}
  \caption{Comparison of the wall-clock times of the different mehods}
  \label{fig:time2d}
\end{figure}

The use of the DNN-MG method in applications is only meaningful when it
not only improves the accuracy compared to a coarse mesh solution
(as demonstrated in the last subsection)
but also reduces the computation time compared to solving on a finer mesh.
As shown in Figure~\ref{fig:time2d}, DNN-MG saves 55\,\% of
the wall clock time required for a fine mesh solution.
It hence indeed improves the computational efficiency.
Compared to solving on a coarse mesh, the runtime increases by 37\,\%,
but with an improved accuracy for DNN-MG.
Fig.~\ref{fig:time2d} also shows that the evaluation of the neural network
requires only 2\,\% of the DNN-MG runtime and
a substantial amount of time is spent on data conversion between
the multigrid framework and the neural network and other auxiliary tasks.
This includes the calculation of the nonlinear residual and the
right hand side, both of which take about as long as the evaluation of the
network. The DNN-MG runtime can hence most likely be further reduced with
a more optimized implementation.

\blue{
\subsection{Prediction across 2 levels}
\label{sec:num:2levels}

So far we only considered the prediction of one level, i.\,e.\ a correction
on level $L$ w.\,r.\,t.\ level $L+1$. In this section we investigate the prediction of a correction using level $L+2$.

The DNN-MG algorithm does not change conceptually in this case. 
Prolongation and restriction, however, now transfer between $L$ and $L+2$. One patch corresponds again to one element on the mesh of level $L$ but this now
includes $4\times 4$ equally subdivided elements of the mesh on
level $L+2$. 
A small modification to the neural network is necessary to accommodate the larger input and output sizes of $99$ and $50$, respectively.
We also increase its internal size to $60\,960$ trainable parameters by setting $N_{\text{GRU}}=96$ to account for the larger number of degrees of freedom.

Figure~\ref{fig:dragliftjmp2} and Table~\ref{tab:draglift} show the
obtained values for the drag and lift functionals.
When comparing to the drag and lift for the $L+1$ case in Fig.~\ref{dl2D_min}, it must be noted that different data is
used for training in the two cases. 
A comparison should thus consider whether the DNN-MG method
is able to learn the respective data (i.e. $L+1$ and $L+2$, respectively) and not whether the total error
can be reduced.  
Figure~\ref{fig:dragliftjmp2} and Table~\ref{tab:draglift} show that with the larger mesh and newtork in the $L+2$ case one obtains predictions for the drag and in particular the lift that are of very high quality
and significantly improved compared to the predictions for $L+1$.
For $L+2$, minimum and maximum of the lift have an error of less than 1\% whereas the lift error was about 15\%
in the $L+1$ case. Further, the enhanced
frequency and amplitude of the lift functional is especially notable.

The computation time of DNN-MG with a two level prediction is very similar to the one level case. 
The additional prolongation and restriction computations do not add noticeable
workload. Copying the larger input and output data and evaluating the larger
network also adds less than $1\%$ to the runtime compared to DNN-MG ($L+1$).
However, we need about 50\% more time to generate the training data for the two level
predictions and the training times increases by $\approx 5\%$.
}

\subsection{Generalizability}
\label{sec:num:generalizability}

For the practicality of the DNN-MG method it is important that a neural network trained on one
flow performs well also on similar ones, i.e. that it generalizes to flows
beyond those seen during training.
The results in Sec.~\ref{sec:flowprediction} already demonstrated that
the network is able to do so under small perturbations
of the geometry of the obstacle.
In the following, we consider three more
substantial changes to the flow: a channel
without an obstacle, a channel
with two obstacles, and a flow in an L-shaped domain,
Unless stated otherwise, we reuse
the network trained as discussed in Sec.~\ref{sec:num:nn:training}
for the channel with a single obstacle.
In this section we again use level $L$ as coarse resolution and $L+1$ as fine one on which the neural network component of DNN-MG operates.


\paragraph{Channel flow without obstacle}

\begin{figure}[t]
  \centering
  \includegraphics[align=t,width=.98\linewidth]{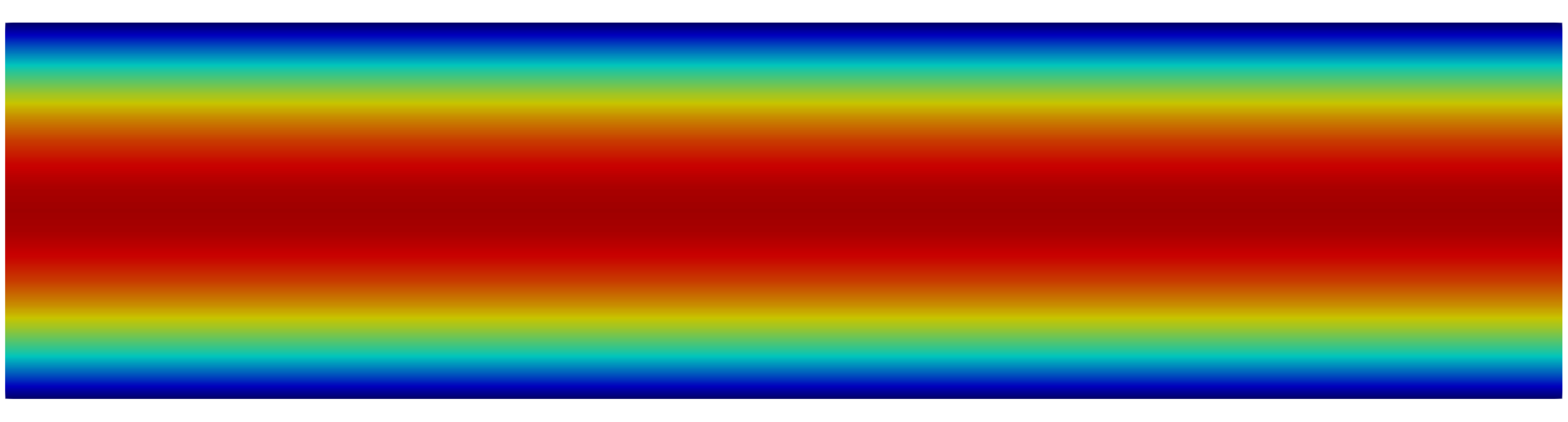}
  \includegraphics[align=t,width=.98\linewidth]{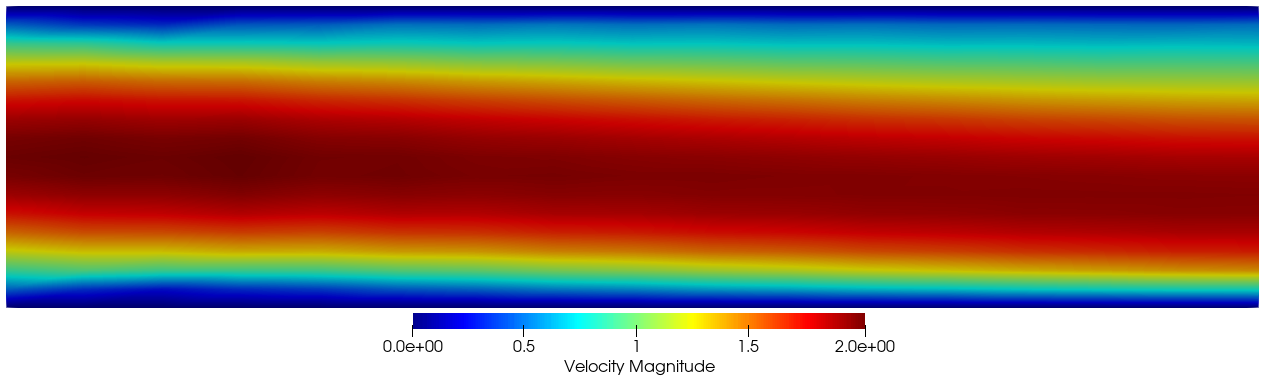}
  \caption{Magnitude of the velocity field obtained with DNN-MG trained on the flow around one cylinder for the channel without obstacle (bottom) and reference solution (top).}
  \label{fig:gen_chan}
\end{figure}

In our first test case we considered the stationary, $1$-dimensional flow that develops in the channel of Fig.~\ref{fig:geom} when the obstacle close to the inflow is removed.
The stationary flow is challenging for DNN-MG in our setup since it is trained on oscillatory flow data only.
Fig.~\ref{fig:gen_chan} shows the DNN-MG solution that has been obtained.
We observe a slight deformation of the velocity field (which should be strictly constant in the horizontal direction) but an otherwise stable and in particular stationary flow.
The network is thus able to generalize to this entirely different flow regime.
We conjecture that the network's ability to do so results from our network design that learns individual patches.
Hence, even for the non-stationary flow around a cylinder that served as training data one has at least locally also relatively stationary flows in the training data.

\paragraph{Channel with two obstacles}

\begin{figure}[t]
  \centering
      \includegraphics[width=\linewidth]{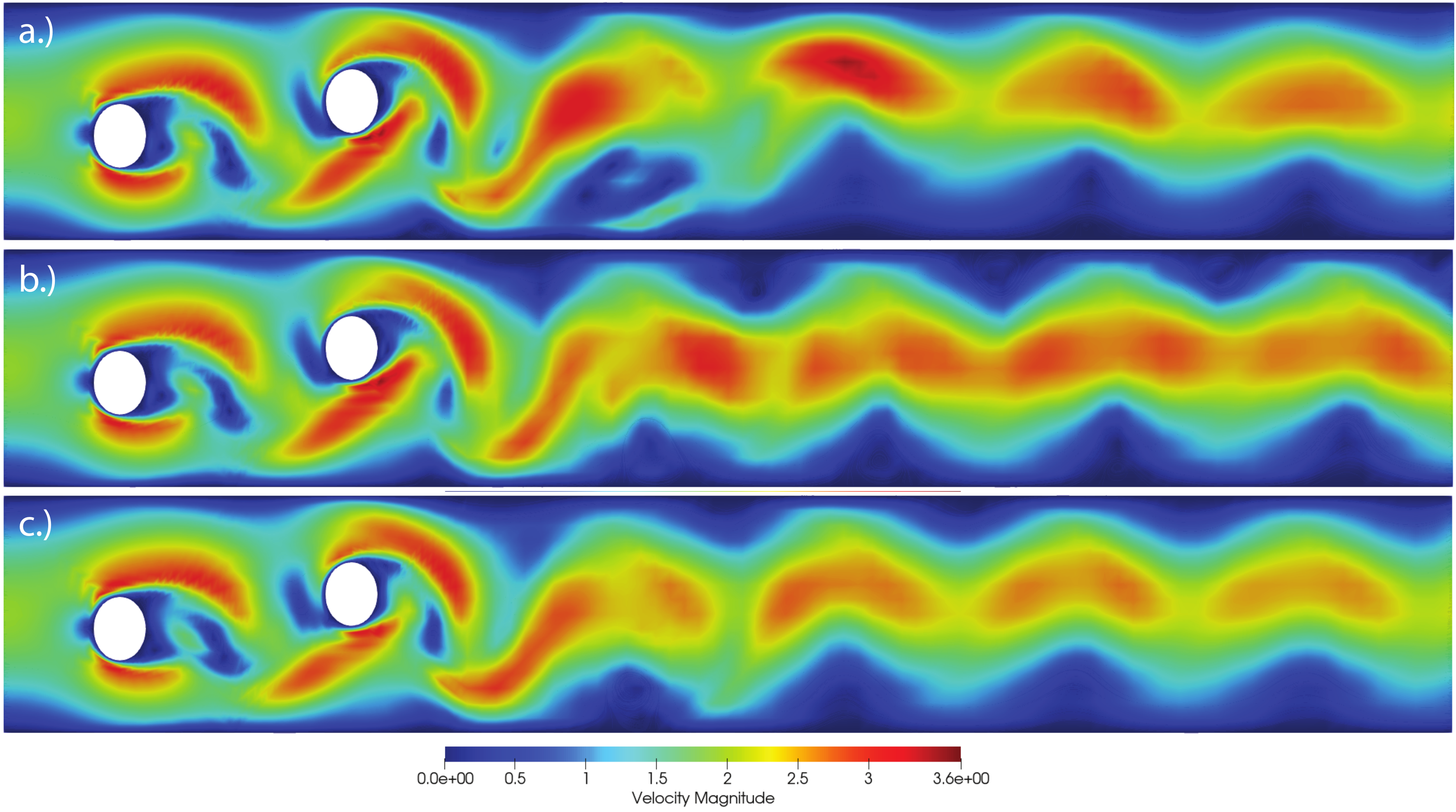}
      \caption{Magnitude of the velocity field of DNN-MG trained on the flow around one cylinder for the channel with two obstacle at time $t=10$ (b.) as well as a multigrid solution with $L+1$ levels (a.) and $L$ levels (c.).}
      \label{fig:velocityfields2obs}
\end{figure}

\begin{figure}[!htb]
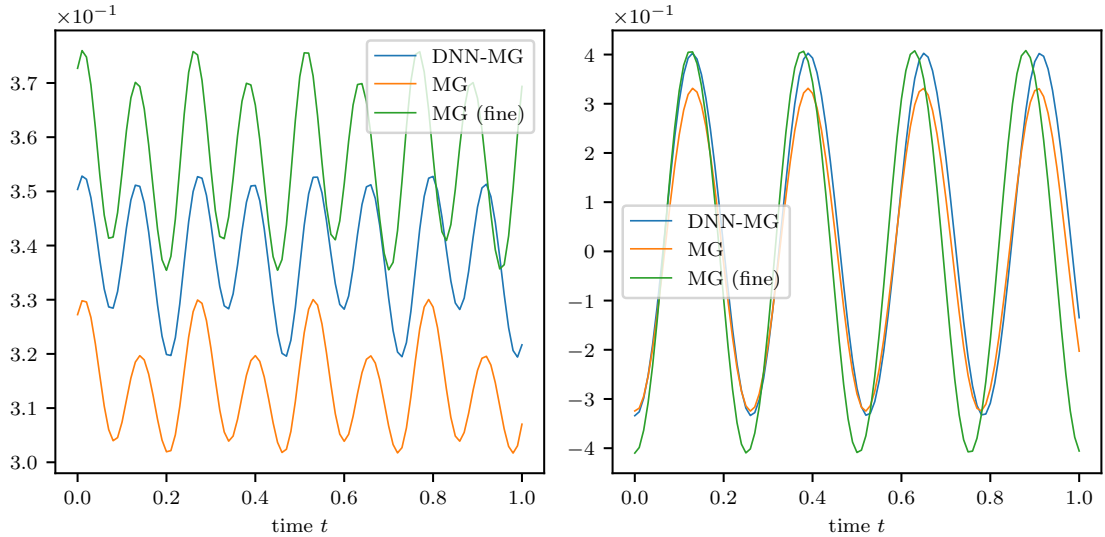

  {\centering
  \resizebox{.49\textwidth}{!}{\input{plots/dragfcts2obs_convminimal.pgf}}
  \resizebox{.49\textwidth}{!}{\input{plots/liftfcts2obs_convminimal.pgf}}}
  \caption{Drag (left) and lift (right) functionals of DNN-MG trained on the flow around one cylinder for the channel with two obstacle as well as for a classical multigrid solution on $L$ and $L+1$ levels.}
  \label{dl2D_min2obs}
\end{figure}

The second generalization test was also an extension of the initial benchmark problem in Sec.~\ref{sec:flowprediction} where we replaced the one obstacle with two. These are centered at $(0{.}2,\,0{.}18)$ and $(0{.}6,\,0{.}24)$, respectively, with their minor and major axes being $0{.}9$ and $1{.}1$.
Since the incoming flow is already disturbed by the first obstacle before reaching the second one, the configuration differs substantially from the training scenario with only one.
To represent the modified geometry of this test case we employ a finite element mesh with $9,741$ degrees of freedom on level $L$ and $37,917$ for the fine mesh of level $L+1$,

The magnitude of the velocity field obtained with DNN-MG as well as multigrid solutions on levels $L$ and $L+1$ are shown in Figure~\ref{fig:velocityfields2obs}.
All fields look similar, although visually DNN-MG is closer to the reference than the coarse solution in the vicinity of the two obstacles and differs more in the far field on the right.
To quantitatively assess the performance of DNN-MG we again used the drag and lift functionals, which we now evaluated for the second obstacle.
Fig.~\ref{dl2D_min2obs} and Table~\ref{tab:draglift:double} show that DNN-MG is able to substantially improve the estimated amplitude, despite being trained with only one obstacle.
Moreover, the results were obtained using a network trained on a different mesh, which was possible since our neural network operates on local patches that are agnostic to the global mesh structure.
The frequency in Fig.~\ref{dl2D_min2obs}  is again suffering from the frequency and phase shift that arises from the different mesh resolutions but that are unrelated to DNN-MG, cf. Sec.~\ref{sec:flowprediction}

\begin{table}[!htb]
  \centering
  \resizebox{\textwidth}{!}{
  \begin{tabular}{l|llll|lllll}
    \toprule
    &\multicolumn{4}{c|}{drag}&\multicolumn{4}{c}{lift}\\
    type (level)            & min    & max    & mean & ampl.& min    & max    & mean & ampl.\\
    \midrule
    MG ($L+1$) & $0{.}335$ & $0{.}375$ & $0{.}355$& $0{.}040$& $-0{.}401$& $0{.}400$& $-0{.}001$& $0{.}802$\\
    DNN-MG ($L+1$) & $0{.}319$ & $0{.}353$ & $0{.}336$& $0{.}034$& $-0{.}303$& $0{.}399$& $ \phantom{-}0{.}048$& $0{.}702$\\
    MG ($L$) & $0{.}302$ & $0{.}330$ & $0{.}315$& $0{.}028$& $-0{.}302$& $0{.}302$& $-0{.}000$& $0{.}602$\\
    \bottomrule
  \end{tabular}}
  \caption{Drag and the lift functionals for the generalization of DNN-MG trained on the flow around one cylinder for the channel with two obstacles. As reference we show the classical multi-grid solutions on the coarse and the fine meshes.}
  \label{tab:draglift:double}
\end{table}

\paragraph{L-shaped domain}

\begin{figure}[t]
  \parbox{.32\linewidth}{
    \includegraphics[width=\linewidth]{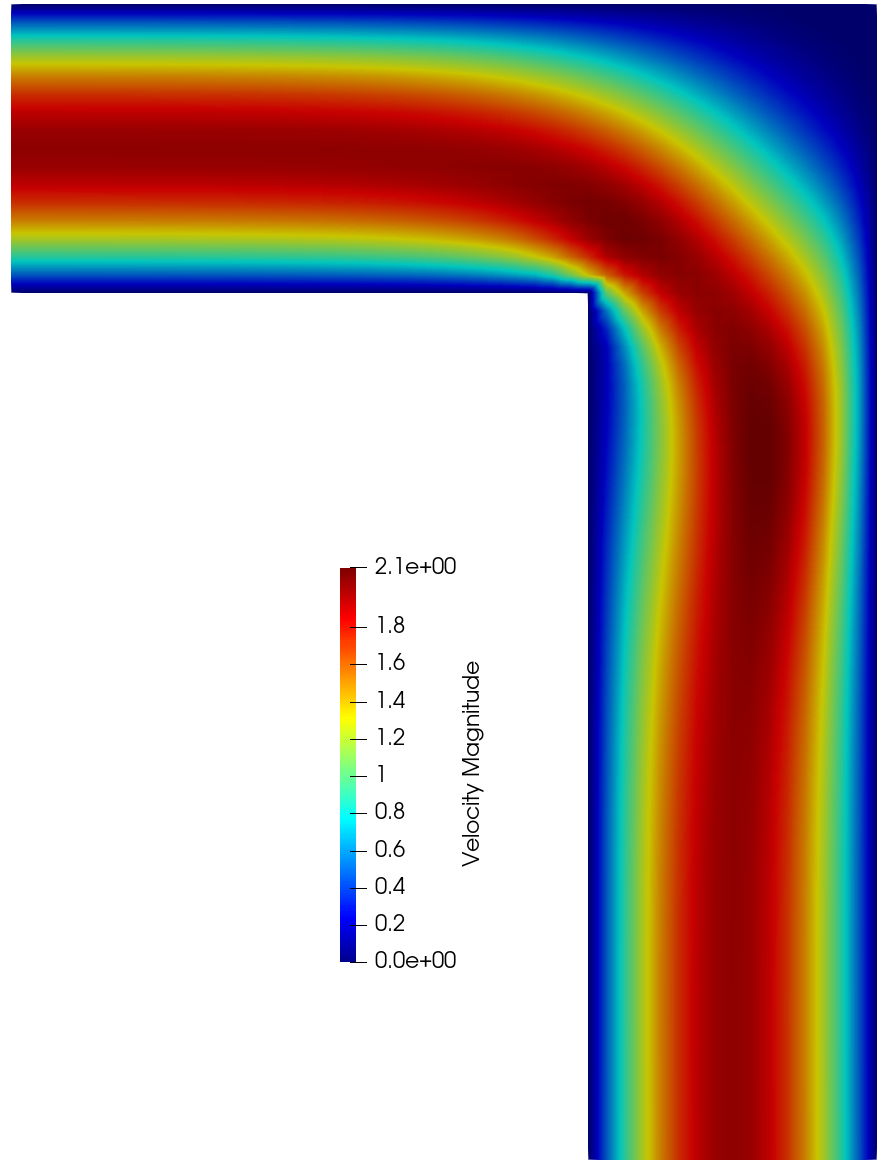}
  }
  \parbox{.32\linewidth}{
    \includegraphics[width=\linewidth]{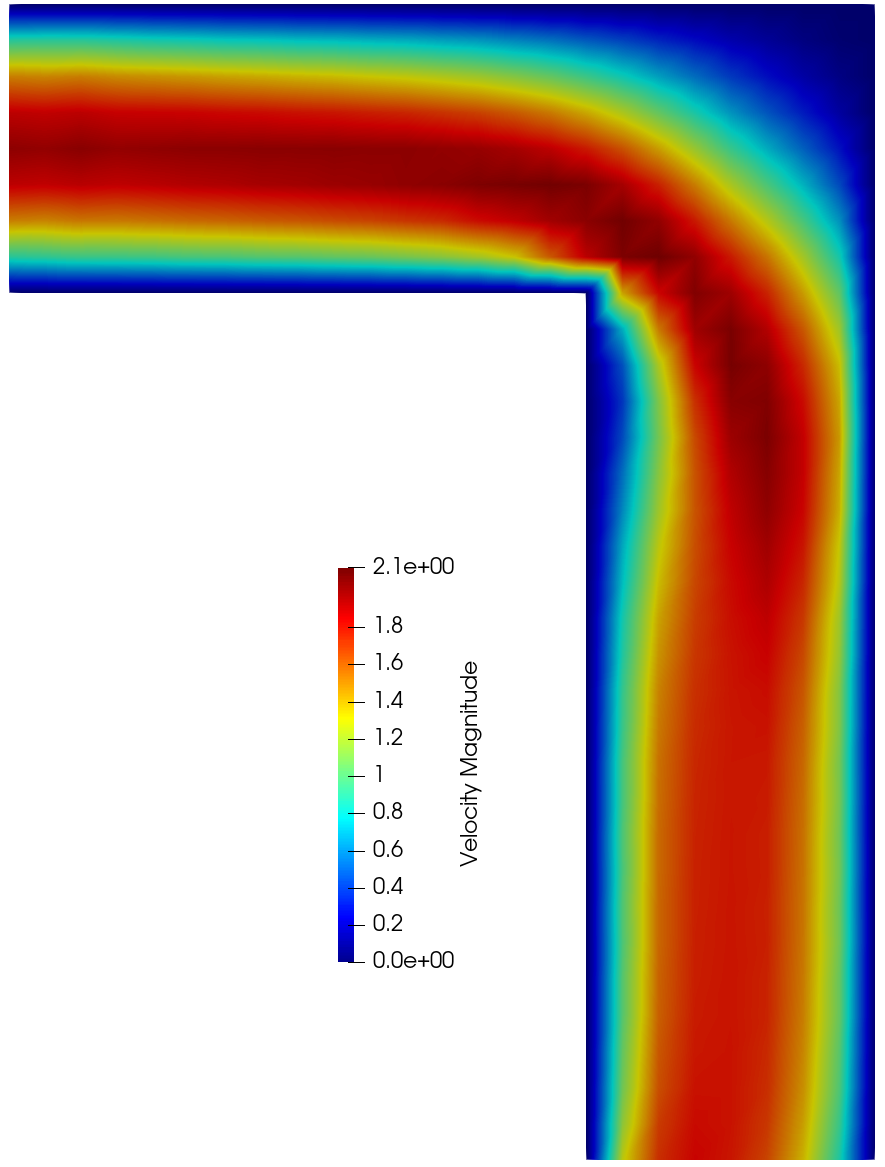}
  }
  \parbox{.32\linewidth}{
    \includegraphics[width=\linewidth]{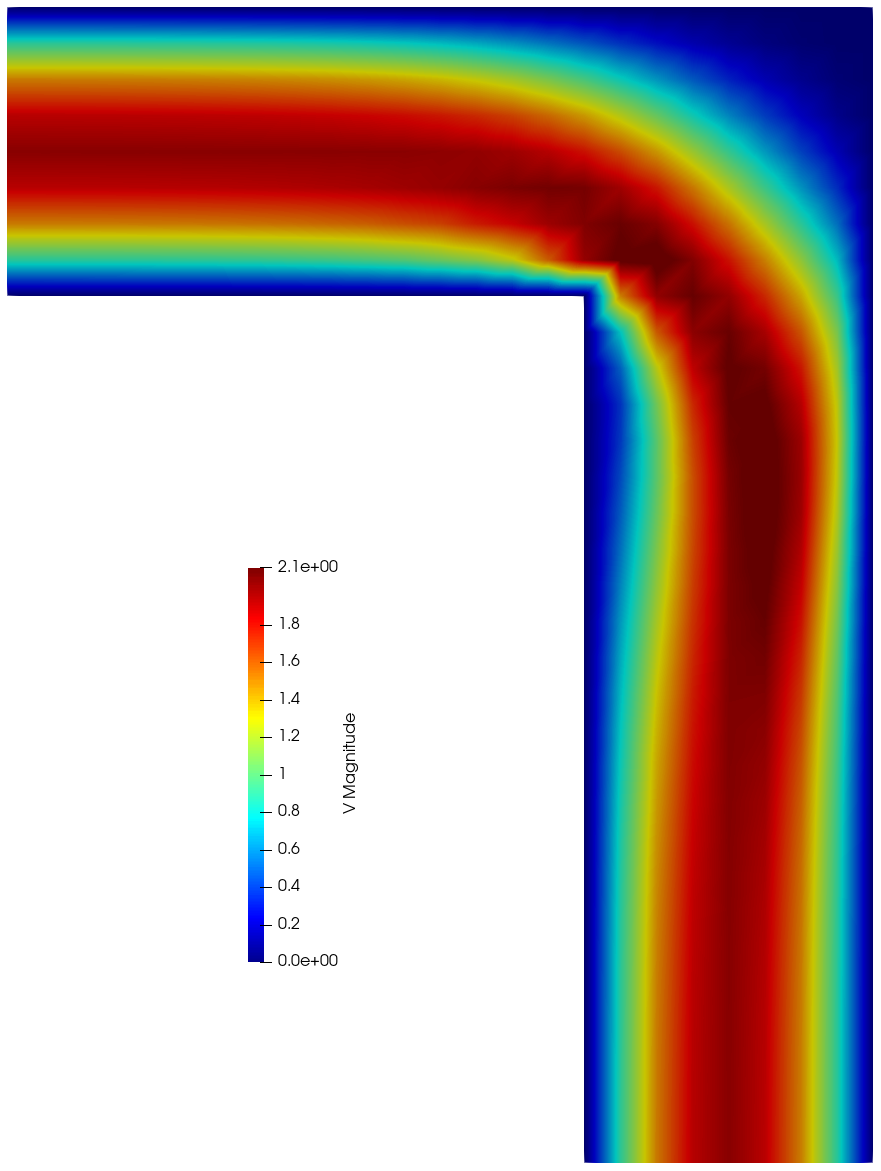}
  }
  \caption{Comparison between simulations with $L+1$ multigrid levels, DNN-MG, $L$ multigrid levels (from left to right).}\label{fig:ldomain}
\end{figure}

As a final test case we considered an L-shaped domain without obstacle as shown in Fig.~\ref{fig:ldomain}.
Using DNN-MG with the trained neural network from Sec.~\ref{sec:flowprediction} fails in this case since the training data contains only flows in the positive $x$-direction (the predicted solution then has a unphysical drift in the $x$-direction also in the lower arm of the L-shape).
To alleviate this problem, we retrained DNN-MG with an augmented data set where the existing data was also used with the $x$- and $y$-axes swapped.
Through this, no new and expensive simulations on level $L+1$ were necessary, and the training converged faster since the optimization could start from an already trained network.
In Figure~\ref{fig:ldomain} the magnitude of the resulting velocity fields for DNN-MG and multigrid solutions on levels $L$ and $L+1$ are depicted.
As can be seen there, DNN-MG does not provide a substantial improvement over the coarse multigrid solution.
In our opinion, this is not surprising given that the training data does not contain a flow past a sharp edge.
Nonetheless, the DNN-MG simulation remains stable and the neural network correction does not deteriorate the solution quality of the coarse simulation on level $L$.

\section{Conclusion}
\label{sec:conclusion}

We have presented the deep neural network multigrid solver (DNN-MG) that uses a recurrent neural network to improve the efficiency of a geometric multigrid solver, e.g. for the simulation of the Navier-Stokes equations.
The neural network is integrated into the multigrid hierarchy and corrects on fine levels the prolongated solutions (instead of the smoothing operations used classically), affecting the overall time evolution through the right hand side of the problem.
Central to DNN-MG is the compactness of its neural network.
This is enabled through the network's locality, namely that it operates independently on small patches of the domain, and the coarse multigrid solution that provides a ``guide'' for the network's corrections. 
The compactness is vital for the computational efficiency of DNN-MG and also reduces the training data and time that are required. 
The locality furthermore facilitates generalizability and even allows one to use a network trained on one mesh domain with an entirely different one.
The use of recurrent neural networks with memory thereby enables DNN-MG to account for time dependencies in a flow and obtain predicted corrections that are temporally coherent.

We analyzed the efficacy of DNN-MG for the classical Navier-Stokes benchmark problem of a laminar flow around an obstacle.
Our results show that DNN-MG is able to substantially improve the accuracy of solutions and provide significantly reduced errors for the drag and lift functionals.
It thereby requires only approximately half the computation time of a multigrid solution at the same resolution and only a modest increase compared to a coarse one.
At the same time, DNN-MG is able to generalize substantially beyond the flow used for training and with the neural network trained on the one obstacle flow we were able to obtain high fidelity solutions also for channel flows with no or two obstacles.

The experiment with the flow in the L-shaped domain, where we do not obtain improvements over a coarse multigrid solution, shows the limits of the current DNN-MG implementation.
We believe that a more complex neural network and more variability in the training data will be able to alleviate the limitation.
We want to study this in future work. 
There, we also want to generalize our current implementation to 3D flows.
Although DNN-MG is already considering the physics of the underlying problem by using the nonlinear residual on level $L+1$, it is worth exploring how additional model knowledge such as the divergence freedom of the predicted velocity fields can be integrated into DNN-MG.
In future work, we also want to consider the application of DNN-MG to other partial differential equations, cf. from nonlinear elasticity, cf. Sec.~\ref{sec:dnnmg:general}.

\paragraph{Acknowledgement}
NM and TR acknowledge the financial support by the Federal Ministry of
Education and Research of Germany, grant number 05M16NMA as well as
the GRK 2297 MathCoRe, funded by the Deutsche Forschungsgemeinschaft,
grant number 314838170.
NM acknowledges support by the Helmholtz-Gesellschaft grant number HIDSS-0002 DASHH.
CL was funded by the Deutsche Forschungsgemeinschaft (DFG, German Research Foundation) – Project-ID 422037413 – TRR 287.

\bibliographystyle{plain}
\bibliography{lit,neuralnets,climate.bib}

\end{document}